\documentstyle[12pt]{article}

\newcommand{\cL}{{\cal L}}
\newcommand{\cS}{{\cal S}}
\newcommand{\nb}{\nabla}
\newcommand{\hf}{\widehat{\varphi}}
\newcommand{\hp}{\widehat{\pi}}

\newcommand{\str}{\stackrel}
\newcommand{\ov}{\overline}
\newcommand{\wh}{\widehat}
\newcommand{\wt}{\widetilde}
\newcommand{\cF}{{\cal F}}
\newcommand{\cH}{{\cal H}}
\newcommand{\cM}{{\cal M}}
\newcommand{\nn}{\nonumber}

\newcommand{\vp}{\varphi}

\newcommand{\Symm}{\mathop{\rm Symm}}
\newcommand{\Asymm}{\mathop{\rm Asymm}}

\newcommand{\Sp}{\mathop{\rm Sp}\nolimits}

\newcommand{\hb}{\hbar}
\newcommand{\bR}{{\bf R}}
\newcommand{\bZ}{{\bf Z}}
\newcommand{\bT}{{\bf T}}

\begin{document}

\title{Quantization of Thermodynamics, Supersecondary Quantization, 
and a New Variational Principle}
\author{V. P. Maslov\thanks{This research was supported by the Russian 
Foundation for Basic Research under grant No.~99-01-01198.}}
\date{}

\maketitle

\begin{abstract}
In solving the problem of finding a temperature distribution
which, at zero temperature, corresponds to superfluidity,
i.e., to nonzero energy, the author tried to quantize free
energy. This was done on the basis of supersecondary
quantization whose special case is the usual secondary
quantization for bosons and with the help of which new
representations of the Schr\"odinger equation were obtained. 
The supersecondary quantization allowed the author to construct 
a variational method whose zero approximation 
are the Hartree--Fock and Bogolyubov--BCSch variational
principles. This method works especially well in the case  
of not a large number of particles.
The new quantization and the variational method
are of general character and can be used in the quantum field 
theory. 
\end{abstract}

\section{Introduction}

\subsection{Different pictures of $N$ bodies and their
representations}

There are two aspects in the many-body problem.
The many bodies can be located in our three-dimensional space
and move in some way.
They can collide or interact.
All this takes place in our three-dimensional space.
However, we can consider these bodies in a different way.
Namely, we can treat these $N$ bodies as one point
(here the term ``$N$ bodies'' means ``$N$ material points'')
in the three-dimensional space.
In this three-dimensional space this single point moves in some
way, and we can observe its motion.
In some cases one of these two approaches (two aspects) is
efficient.
In other cases the other approach can be useful.
So, from the mathematical viewpoint,
it is simpler to deal with a single point.
{}From the physical viewpoint, since we live in the
three-dimensional space, we would like to understand how
material points behave in our three-dimensional space.
This is a more natural ``picture.''
{}For example, in the three-dimensional space we can study a
particle located in the field of all other particles.
In particular, this approach underlies Vlasov's theory of
self-consistent field. In this theory it is assumed that all
particles are approximately the same and have the same
distribution.
We can assume that one particle behaves as if it interacts with
another particle, and this other particle is the same particle,
i.e., it has the same distribution as the first particle.

It turns out that this approach can be realized not only
approximately, as it was realized by Vlasov,
but exactly by using the secondary quantization, i.e.,
in other words,
in the three-dimensional space we consider
the behavior of operators rather than the behavior
of functions.
Namely, we study some creation and annihilation operators.
{}For the first time, this was realized by Dirac in quantum
mechanics.
However, it is not necessary to consider just the quantum
mechanics. So, Schoenberg studied the classical mechanics
(approximately in 1953--1956) and applied this secondary
quantization method to classical objects and to classical
statistical physics.

Indeed, this is a very general thing.
{}For example, I together with my pupil used this method in the
case of $N$ fields. We acted as if it was the third
quantization, which, in fact, is a version of the secondary
quantization [1].
Here we have a natural generalization. If we consider $N$
particles as a single particle in the three-dimensional space,
why cannot we consider, for example, any two particles of these
$N$ particles?
I can treat this pair as one particle in the six-dimensional space
and then study these pairs of six-dimensional particles.
Each of them is similar to one particle in the
$3N/2$-dimensional space. So I can simultaneously consider a
pair of particles and a single particle.
In other words,
there exist some six-dimensional particles (pairs)
and some three-dimensional particles
and all this is embedded in the corresponding space. If
there are $k$ pairs and $m$ ``single'' particles, then
we have $2k+m=3N$, where $N$ is the general number of particles.

In fact, in addition to the coordinate of the particle
(the location of the point), there can be some other degrees of
freedom.
{}For example, in the classical mechanics there can be momenta.
In the quantum theory, there can be spins and some other degrees
of freedom.

In the theory developed here, I introduce one more degree of
freedom, the minimal degree which is just a number.
In addition to the location of a particle in some part of
the space, this particle is equipped with a number.
This viewpoint had a good interpretation in the early quantum
mechanics.
In the Bohr theory we deal with electrons.
We can interchange their places, but nothing changes.
This is the identity principle.
However, besides these electrons, there are orbit numbers.
They can also be treated as numbers assigned to electrons.

{}For example, I have $N$ particles and I consider some pairs of
particles and single particles, which I have numbered.
Perhaps, it is convenient also to number the pairs.
Whether I have numbered then or not, this does not change
anything, since this numbering is not related to the fact
that I fix these particles and they cannot interchange their
places.
Thus all this is the same picture. These are simply different
representations of the same phenomenon.
The secondary quantization allows us to write equations in the
three-dimensional space but with operators, i.e.,
an operator equation.
This operator equation is equivalent to one equation in the
$3n$-dimensional space (which is simply an equation)
not for operators but for functions.
The same problem is obtained in terms of
representations, which I propose and call supersecondary
quantized.

\subsection{Series and serials}

Probabilities of transition from level to level play an
essential role.
In particular, as is known, the energy quantums issued by an atom
correspond to transitions from level to level.
If the probability of transition is zero or very small,
i.e.,
if this transition is practically forbidden,
then the corresponding levels are treated as if they belong to
different series.
In fact,
the partition into series means a partition into
systems that are isolated from each other.
If a given series of energy levels contains the minimal level,
which is the principal state of the series,
then this is a metastable state.
If the system is in this state and transitions to the lower
levels of other series are almost forbidden,
then the system stays in this state sufficiently long.

The existence of a superfluid liquid or a superconducting
current corresponds to such metastable series.
The current can
flow without friction during 100000 years, which means that the
series is almost stable.
In a finite large volume, where we consider a system with a
large number of particles, the velocities and currents
are discrete, and each value determines its own metastable
series.
A set of series that differ in velocities we call a {\it serial}.

To each picture discussed above, there corresponds its own
serial in the asymptotic approximation.
Although the physicists do not say this directly,
in reality, they try to find a serial such that
the lower level of its lower series
(i.e., the level at the zero velocity)
would coincide with the principal state of the entire original system.

In the above-mentioned paper by N.~N.~Bogolyubov (1947),
such a serial was found. It corresponds to the following
picture: $N$ particles are in the three-dimensional space and
all their numbers coincide.

However, it turns out and this will be proved in the present
work that the serial corresponding to the picture in which
pairs of particles are not numbered but the particles are
numbered has a level which is lower than that of the
Bogolyubov serial. In this sense, apparently, the serial
corresponds to the true phenomenon of superfluidity and, to a
larger extent, corresponds to the experiment.
Although, if the nature ``set'' a system to another serial, the
system will stay in this state for a long time.

However, it is possible that resonances may occur between
different series. In this case the series is destroyed.
An example is provided by the one-dimensional Schr\"odinger
equation with two minima.
If the maximum between the minima is
very large and if we have a semiclassical approximation,
then
to each hollow there corresponds its own series and
if there are no resonances, then transitions
>from one hollow to the other are almost forbidden
(the ``tunnel'' effects).
If some eigenvalues of two series coincide, then the resonance
is possible and in this case the position of a particle in one
of the hollows at this eigenvalue is destroyed.
{}For example, in the potential symmetric with respect to the maximum,
all eigenvalues ``almost'' coincide, all are in resonance, and
the system does not split into series.

\subsection{Quantization of free energy}

We assume that the eigenfunctions of the Schr\"odinger equation
(the Hamilton operator) remain the same.
{}From each eigenvalue we subtract the temperature multiplied by
entropy corresponding to this eigenvalue.
In the simplest
case of a discrete spectrum, as is known~[2],
this is the logarithm of the multiplicity of the eigenvalue.

As is known, the Hamilton operator can be represented as the
sum of its eigenvalues multiplied by the projectors on the
corresponding subspaces of eigenfunctions.
The free energy operator,
as was already pointed out, is the sum of the same
eigenvalues
>from which the temperature multiplied by entropy is
subtracted
and then they are multiplied by the same projectors.
If we consider the principal state of this quantum free energy, then
we obtain all the results of the former thermodynamics.
But we can consider not only the lower level of this serial but
also the series corresponding to the velocities (currents).
Then we obtain answers to the question of how the metastable states
vary with temperature.

We shall show that there exist serials that contain superfluid
and superconducting series at any temperature.
However, these series necessarily do not contain the principal
state, i.e., there exist serials whose lower level is lower.
By idea, the nature must choose the latter.
On the other hand, by the argumentation of the great Dirac,
``it were astonishing if the nature did not use this possibility.''
I add ``somewhere'', in nuclei, in stars, etc.
in one word, somewhere.

Is it possible to create such serials ``by hand''?
We have already created by hand metastable series, higher than
the principal state, by letting currents flow.
Is it possible, by using resonance, also to ``pump''
high-temperature serials?

\subsection{Necessary conditions for metastable states}

If a large parameter is given as some concrete number,
it is sometimes not clear whether this parameter is ``large''
or ``not very large.''
Say, if the number of particles is 100, this number must be
a sufficiently large parameter, but the logarithm of this large
parameter is already not too large.
In this case, first of all, we would like to know how to
determine a metastable state.
I introduce, at least, necessary conditions for a metastable
state.

To this end, I first consider a very simple example, the
one-dimensional Schr\"odinger equation whose potential well has
two hollows, i.e., there are two minima: one minimum is global,
i.e., it corresponds to the principal state, and the other minimum is
local. In the case of the classical asymptotics, it determines a
metastable state. What does it mean?

If the system is at this lowest level corresponding to the local
minimum or to the smaller hollow,
i.e., in other words, if the eigenfunction is concentrated near the
local minimum and rapidly decreases everywhere else,
then the following question arises: what happens if the system
if perturbed?
In other words, what is the probability of the transition
>from this state to the principal state, i.e., to the state
that corresponds to the global minimum and whose eigenfunction
is concentrated near this minimum and decreases sufficiently
rapidly everywhere else.

If we have a semiclassical approximation, then the tails of
these eigenfunctions decrease very rapidly.
Hence if we consider
the matrix element obtained as a result of perturbation
by some operator of multiplication by an external potential,
then we see that this matrix element is very small, since the
product of functions is very small.
In other words, I take the following matrix element:
$\psi_0$ is the eigenfunction corresponding to the principal state
and $\psi_1$ is the first eigenfunction corresponding to the
other hollow.
$V'$ is the potential by which we perturb the system.
If the matrix element $(\psi_0,V'\psi^*_1)$ is small,
then this means that the transition probability is small.
Hence this state is metastable, i.e.,
it is very difficult to ``fall'' to the
lower level, while transitions to the higher levels
whose supports are also
near the minimum of this smaller hollow are quite possible.

However, since we say that this metastable state is, to some
extent, a model of superfluidity and superconductivity,
it is very important for us that the energy did not become less.
If the energy increases, then the larger the better.
(If the velocity increases due to this perturbation, this is
also very good.)
So, we perturbed the problem as if by friction, as Landau would say,
and nevertheless, the energy did not decrease, since the
transition to lower levels for such an asymptotic problem is
practically forbidden.

However, if we take an operator not from the class I described
but from a more complicated class, e.g., from the class where
the translation operator is also involved,
then the matrix element is not at all small.
Hence the first problem is to specify the operators by which we
perturb our problem.

The operators by which we perturb the problem are
operators that preserve the support of a function,
i.e., in other words,
if a function $\psi$ is sufficiently smooth and equal to zero
outside some interval (thus is interval is called just the
support of this function) and we apply the multiplication
operator or the differentiation operator of any order, then
these operators do not take the function outside its support.
We usually perturb our problem by such operators.
What does it mean?

This means that, prior saying that we have a metastable state,
we must define the class of operators with respect to which this
state is metastable and for which the transition operators are
small.
Natural operators of this type are just the potential forces.
{}Further, the question arises:
perhaps, this is not a semiclassical approximation,
the parameter $\hb$ is not small, perhaps, the barrier is not
very high, what happens then?
Can a metastable state exist in this situation and how long?
In this section we shall try to answer this question.

The first question: is such a state possible at all or not?
The second question: how small are the matrix elements of the
transition?
Thus the first question is reduced to the following one:
does the minimum exist, or not, and of what is this minimum?

We spoke about the minimum of a potential well.
However, about what minimum must we speak in the quantum
problem itself, without a small parameter?

In the quantum problem without a small parameter we can reason
in the following way: we consider a set of perturbing operators
with support on the interval $[a,b]$, in other words,
we take the operators of multiplication by smooth finite functions
that differ from zero on the interval $[a,b]$.
Then we act by the set of such operators on the entire space
$L_2$.
We obtain the set of values of the result of this action.
We study whether the original Hamiltonian attains its local
minimum on the functions from this set.
Namely, this minimum is understood as the true local minimum
with positive second-order derivatives, i.e.,
the minimum does not lie on the ``edge'' of the interval but
lies somewhere in the middle.
If the interval $[a,b]$ includes the minimum point
of our potential, i.e., the smaller hollow,
then the local minimum necessarily exists for a sufficiently
small $\hbar$.
Such a minimum can also exist if $\hbar$ is not very small,
but just the existence of such a minimum answers our question of
whether there exist a metastable state.

If there is no minimum, we cannot speak about a metastable
state. If there is no second hollow, then, apparently, the
Hamiltonian does not have a minimum, at least for sufficiently
small $\hbar$.

At all events, we, first of all, must answer the question
of whether the quantum problem with classical Hamiltonian, which
is not related to the classics, has a local minimum.
If such a local minimum exists, then there exists an eigenvalue,
which is the nearest to this minimum, and the corresponding
eigenfunction determines some metastable state of the original
operator.
In this case a small perturbation of the length of the interval
does not lead to the transition to another value.

The value of the dimensionless variable $V_0a^2m/\hbar^2$ at
which the local (not principal) minimum disappears will be called
the {\it critical value}.
Here $m$ is the mass, $\hbar$ is the Planck constant, and the
potential has the form $V_0 v(x/a)$, where $V_0$ is a constant,
$a$ is the characteristic length, and $v(y)$ is the potential
well with two hollows mentioned above.

The question of how long this metastable state lives is still
open, hence I only say that this condition is necessary for the
existence of a metastable state.

\subsection{Necessary conditions for the existence of
superfluidity and superconductivity}

Obviously, a metastable state does not depend on the
representation. However, we relate the problem of finding a
minimum to some asymptotic problem,
where the representation in which we seek the minimum
(namely, the $k$-representation) is chosen in some natural way.
Similarly, we act in the many-body problem, in the situation in
which the number of particles is not very large.
We first determine the representation for which these metastable
states are natural in our asymptotic problem.

Hence in the many-body problem we choose the representation in
which, as $N\to\infty$ we already have series with metastable
lower levels. The most convenient way is to deal with the
representation of occupation numbers, which is introduced in the
supersecondary quantization, and hence, in the variational
method, which will be considered later.

These occupation numbers $n_j$ attain integer nonnegative values
for $j=0,1,2,,\dots$.

Let us consider the set of functions of occupation numbers
$A(n_0,n_1,n_2,\dots)$ such that
$$
\sum_{\{n\}}|A(n_0,n_1,n_2,\dots)|^2<\infty
$$
and $A(n_0,n_1,n_2,\dots)=0$ if $N\ne n_0+n_1+n_2+\dots$.
The subset of functions of occupation numbers such that
$A(n_0,n_1,n_2,\dots)=0$ for $\alpha N< n_0+n_1+n_2+\dots$
will be the {\it characteristic class\/} of functions
corresponding to the number $0<\alpha<1$.

To determine a metastable state in the problem of $N$ bodies,
we seek the minimum of the Hamiltonian in the representation of
occupation numbers on the set of functions from the characteristic
class.
By analogy with the preceding section,
we can find the critical value $N_{\rm cr}$ for which
the metastable state disappears.
Superfluidity and superconductivity in the system are possible
for $N$ larger than the critical values.
Thus, $N_{\rm cr}$ is the lower bound for $N$ for which
superfluidity and superconductivity are possible at least for a
short time.

\subsection{Variational methods and variational principles}

There exist many variational methods,
for example, the variational Riesz method, the variational
Galerkin method, the variational Tamm--Dankov method.
The method of finding the solution is, so to say, successive,
i.e.,
there {\it a priori\/} exists a sequence that converges
to the solution.
In the Tamm--Dankov method this convergence is not proved, but
each successive approximation must be, in principle, better than
the preceding approximation.

The situation with variational methods is somewhat different.
The variational methods are formulated so that they directly
lead to some equation and are postulates that follow from the
equations.
There also exist variational principles that are derived from
the original variational principle which is considered
on a narrower function class. They lead to some new equations.
The relation between these new equations and the original
``true'' equations is determined experimentally.

Suppose that some variational method, say, the Hartree--Fock
method is used. The results obtained are compared with
experimental results, and the coincidence is good.
In the variational principles such as
the Hartree--Fock principle,
the variational Bogolyubov principle,
the variational Schriffer principle,
we try to guess the form of the solution by substituting this
solution and obtaining some new equation.
So, these principles lead to some new equations
and, instead of the original equations,
we solve some other equations obtained from the variational
principle.
We could determine the relation between
the variational principle and the equations
if we had a sequence
(as in the Riesz method, in the Galerkin method, and in some
other methods) of test functions with the help of which we could
step-by-step approach the solution,
but not a single test function with the
help of which we need to verify the variational principle.

Just in this case,
the supersecondary quantization, which I propose,
naturally leads to a variational method, i.e.,
to the method of successive approximations.
To some extent this ideology
is close to that of the Tamm--Dankov method.
The variational Bogolyubov and Hartree--Fock principles
constitute only the first approximation.
For some problems this yields exact asymptotic methods, as was
shown by Bogolyubov for the Cooper--Bardeen model.
He showed that this is an exact asymptotics as $N\to\infty$.
It is of interest that the variational method proposed here
leads to better results if the number of particles is not very
large, just where the asymptotic theory cannot be applied.

If the number of particles is 104, as in traps,
then this case cannot be an asymptotics as $N\to\infty$, since
the logarithm, which tends to infinity as $N\to\infty$, is here
equal to~$4$.
If we have $10^{23}$, just as in the weakly nonideal Bose gas,
then the logarithm is already a sufficiently large number, and
this case can be an asymptotics for large $N$.
In this case it is not so convenient to use the variational
method.

\section{Supersecondary quantization in the boson case}

{}Following [3], we introduce the averaging of operators
in the more general case of a large number of two types of clusters. Let
$k\ge1$ and $N_1,\dots,N_k\ge0$ be integers. Consider the
Hilbert space of functions of the form
\begin{eqnarray}
&&\Psi\big(x^1_1,j^1_1;\dots;x^1_{N_1},j^1_{N_1};x^2_1,x^2_2,j^2_1;
\dots;x^2_{2N_2-1},x^2_{2N_2},j^2_{N_2};\dots\nn\\
&&\qquad x^k_1,\dots,x^k_k,j^k_1;\dots;
x^k_{kN_k-k+1},\dots,x^k_{kN_k},j^k_{N_k}\big),
\end{eqnarray}
where $x^l_{p_l}\in \bT^3$, $\bT^3$ is the
three-dimensional torus with side length $L$,
$j^{l}_{q_l}=1,\dots,\infty$
for all $l=1,\dots,k$ and $p_l=1,\dots,lN_l$, $q_l=1,\dots,N_l$.
moreover,
the functions (\ref{f18}) are symmetric with respect to the
permutations of any pairs
$(x^{l}_{pl-l+1},\dots,x^{l}_{pl},j^{l}_{p})$ and
$(x^{l}_{ql-l+1},\dots,x^{l}_{ql},j^{l}_{q})$
of sets of variables, where
$l=1,\dots,k$ and $p,q=1,\dots,N_l$.

The inner product has the form
\begin{eqnarray*}
&&(\Psi,\Phi)\str{\rm def}{=}
\sum^{\infty}_{j^1_1=1}\dots\sum^{\infty}_{j^1_{N_1}=1}\dots
\sum^{\infty}_{j^k_1=1}\dots\sum^{\infty}_{j^k_{N_k}=1}
\int\dots\int dx^1_1\dots dx^1_{N_1}\dots dx^k_1\dots dx^k_{kN_k}\times\\
&&\qquad
\times
\Psi^*\big(x^1_1,j^1_1;\dots;x^1_{N_1},j^1_{N_1};
\dots;
x^k_1,\dots,x^k_k,j^k_1;
\dots;
x^k_{kN_k-k+1},\dots,x^k_{kN_k},j^k_{N_k}\big)\\
&&\qquad
\times
\Phi\big(x^1_1,j^1_1;\dots;x^1_{N_1},j^1_{N_1};
\dots;
x^k_1,\dots,x^k_k,j^k_1;
\dots;
x^k_{kN_k-k+1},\dots,x^k_{kN_k},j^k_{N_k}\big)
\end{eqnarray*}

We denote this space by $\cF_{N_1,\dots,N_k}$.
Let us define the boson Fock space $\cF^B_{N_1,\dots,N_k}$.

\hbox{}

{\bf Definition}.
The space $\cF^B_{N_1,\dots,N_k}$ is the subspace of
$\cF_{N_1,\dots,N_k}$ formed by the elements symmetric
with respect to arbitrary permutations
$x^l_p\leftrightarrow x^m_q$ for $l,m=1,\dots,k$ and
$p=1,\dots,lN_l$, $q=1,\dots,mN_m$.
The projection on
this subspace will be denoted by $\wh{P}^B_{N_1,\dots,N_k}$:
$$
\wh{P}^B_{N_1,\dots,N_k}:\,\, \cF_{N_1,\dots,N_k}\to \cF^B_{N_1,\dots,N_k}.
$$

\hbox{}

Now consider the Hilbert space $\cF$ whose elements are
infinite sequences
$$
\Psi=\{\Psi_{N_1,\dots,N_k}\},\qquad
\Psi_{N_1,\dots,N_k}\in\cF_{N_1,\dots,N_k},\qquad N_1,\dots,N_k = 0,1,\dots,
$$
$$
\|\Psi\|^2=\sum^{\infty}_{N_1=0}\dots\sum^{\infty}_{N_k=0}
\|\Psi_{N_1,\dots,N_k}\|^2<\infty.
$$
This space is the infinite direct sum
$$
\cF=\bigoplus_{N_1,\dots,N_k}\cF_{N_1,\dots,N_k}
$$
and is a special case of the Fock spaces [4].

In this space  we introduce the creation operators
${\wh{b}}^+_l(x_1,\dots,x_l,j)$
and the annihilation operators
${\wh{b}}^-_l(x_1,\dots,x_l,j)$, $l=1,\dots,k$.

Let $\wh{P}_{N_1,\dots,N_k}$ be the projection on the
corresponding component of the direct sum,
$$
\wh{P}_{N_1,\dots,N_k}:\,\cF\to\cF_{N_1,\dots,N_k},\qquad
\wh{P}_{N_1,\dots,N_k}\Psi=\Psi_{N_1,\dots,N_k},
$$
and let $\wh{i}_{N_1,\dots,N_k}$ be the operator
of embedding of $\cF_{N_1,\dots,N_k}$ in $\cF$:
$$
\wh{i}_{N_1,\dots,N_k}:\,\cF_{N_1,\dots,N_k}\to\cF,\qquad
\wh{i}_{N_1,\dots,N_k}\varphi
=\{\delta_{N_1M_1}\cdot\dots\cdot\delta_{N_kM_k}\varphi\},
$$
where $\delta_{NM}$ is the Kronecker delta.

We introduce the notion of an averaged operator. Let
$\{\wh{A}_{N}\}$ be a sequence of operators in the subspaces of
$L^{2} (\underbrace{\bT^3 \times \dots \times \bT^3}_{N \mbox{ times}})$
formed by elements $\psi (x_1, \dots, x_N)$
symmetric with respect to the permutations of the
variables $x_j$ and $x_k$, $x_j\in \bT^3$.
To these operators, we assign the operators
$\wt{\wh{A}}_{N_1,\dots,N_k}$, $N_1+2N_2+\dots+kN_k=N$
defined on the subspaces $\cF^B_{N_1,\dots,N_k}$
and acting as the identity operators with
respect to the discrete variables~$j^l_p$.

\hbox{}

{\bf Definition}. The {\it averaged operator} $\ov{\wh{A}}$
is the operator
$$
\ov{\wh{A}}
=\sum^{\infty}_{N_1=0}\dots\sum^{\infty}_{N_k=0}
(N_1+\dots+kN_k)!
\wh{i}_{N_1,\dots,N_k}
\wt{\wh{A}}_{N_1+\dots+kN_k}
\wh{P}^B_{N_1,\dots,N_k}\wh{P}_{N_1,\dots,N_k}
$$
in the Fock space $\cF$.

\hbox{}

{\bf Remark}.
The operator
$\wh{i}_{N_1,\dots,N_k}
\wt{\wh{A}}_{N_1+\dots+kN_k}
\wh{P}^B_{N_1,\dots,N_k}\wh{P}_{N_1,\dots,N_k}$
represented as
\begin{eqnarray}
&&
\frac1{N_1!\dots N_k!}
\sum^{\infty}_{j^1_1=1}\dots\sum^{\infty}_{j^1_{N_1}=1}
\dots\sum^{\infty}_{j^k_1=1}\dots\sum^{\infty}_{j^k_{N_k}=1}
\int\dots\int
dx^1_1\dots dx^1_{N_1}\dots dx^k_1\dots dx^k_{kN_k}
\nn\\
&&\qquad
\times
\stackrel{2}{\wh{b}^+_1}(x^1_1,j^1_1)
\dots
\stackrel{2}{\wh{b}^+_1}(x^1_{N_1},j^1_{N_1})
\dots
\stackrel{2}{\wh{b}^+_k}(x^k_1,\dots,x^k_k,j^k_k)
\dots
\nn\\
&&\qquad
\times
\stackrel{2}{\wh{b}^+_k}(x^k_{kN_k-k+1},\dots,x^k_{kN_k},j^k_{N_k})
\wh{A}_{N_1+\dots+kN_k}
\nn\\
&&\qquad
\times
\Symm_{x^1_1\dots x^1_{N_1}\dots x^k_1\dots x^k_{kN_k}}
\Big\{
\stackrel{1}{\wh{b}^-_1}(x^1_1,j^1_1)
\dots
\stackrel{1}{\wh{b}^-_1}(x^1_{N_1},j^1_{N_1})
\dots
\nn\\
&&\qquad\qquad\qquad\qquad\qquad
\times
\stackrel{1}{\wh{b}^-_k}(x^k_1,\dots,x^k_k,j^k_k)
\dots
\stackrel{1}{\wh{b}^-_k}(x^k_{kN_k-k+1},\dots,x^k_{kN_k},j^k_{N_k})
\Big\}
\nn\\
&&\qquad
\times
\exp\Big(
-\sum^{\infty}_{j=1}\int dz\,
\stackrel{2}{\wh{b}^+_1}(z,j)\stackrel{1}{\wh{b}^-_1}(z,j)
-\dots
\nn\\
&&\qquad\qquad\qquad
-\sum^{\infty}_{j=1}\int\dots\int dz_1\dots dz_k\,
\stackrel{2}{\wh{b}^+_k}(z_1,\dots,z_k,j)
\stackrel{1}{\wh{b}^-_k}(z_1,\dots,z_k,j)
\Big),
\label{f18}
\end{eqnarray}
where $\Symm_{x_1\dots x_N}$
is the operator of symmetrization
with respect to the variables $x_1,\dots,x_N$ and numbers over
operators indicate the order of action of these operators
[5].
The expression (\ref{f18}) was obtained in [6]
for the special case $k=2$ and can be proved in a similar way
for arbitrary $k$.

\hbox{}

According to this remark, the following functional will be
called the {\it symbol} of the averaged operator $\ov{\wh{A}}$:
\begin{eqnarray}
&&A(b_1^*(x_1^1,j^1),b_1(x_1^1,j^1),\dots,b_k^*(x_1^k,\dots,x_k^k,j^k),
b_k(x_1^k,\dots,x_k^k,j^k))=\nn\\
&&\qquad=\sum_{N_1=0}^{\infty}\dots\sum_{N_k=0}^{\infty}
\frac{(N_1+2N_2+\dots+kN_k)!}{N_1!N_2!\dots N_k!}\times\nn\\
&&\qquad\times\sum^{\infty}_{j^1_1=1}\dots\sum^{\infty}_{j^1_{N_1}=1}
\dots\sum^{\infty}_{j^k_1=1}\dots\sum^{\infty}_{j^k_{N_k}=1}
\int\dots\int
dx^1_1\dots dx^1_{N_1}\dots dx^k_1\dots dx^k_{kN_k}\times\nn\\
&&\qquad\times b^*_1(x^1_1,j^1_1)\dots b^*_1(x^1_{N_1},j^1_{N_1})
\dots b^*_k(x^k_1,\dots,x^k_k,j^k_k)\dots
b^*_k(x^k_{kN_k-k+1},\dots,x^k_{kN_k},j^k_{N_k})\times\nn\\
&&\qquad\times\wh{A}_{N_1+\dots+kN_k}
\Symm_{x^1_1\dots x^1_{N_1}\dots x^k_1\dots x^k_{kN_k}}
\Big\{ b_1(x^1_1,j^1_1)\dots b_1(x^1_{N_1},j^1_{N_1})\dots\times\nn\\
&&\qquad\times b_k(x^k_1,\dots,x^k_k,j^k_k)\dots
b_k(x^k_{kN_k-k+1},\dots,x^k_{kN_k},j^k_{N_k})\Big\}\nn\\
&&\qquad \times\exp\Big(-\sum^{\infty}_{j=1}\int dz\,
b^*_1(z,j)b_1(z,j)-\dots-\nn\\
&&\qquad-\sum^{\infty}_{j=1}\int\dots\int dz_1\dots dz_k\,
b^*_k(z_1,\dots,z_k,j)b_k(z_1,\dots,z_k,j)\Big),
\end{eqnarray}
where
$b_l(\cdot,\dots,\cdot,j^l)\in L^2(\bT^{3})$, $l=1,\dots,k$,
for all $j^l=0,1,\dots$ and the inequality
$$
\sum_{j=0}^{\infty}\int\dots\int dx_1\dots dx_l
b_l^*(x_1,\dots,x_l,j)
b_l(x_1,\dots,x_l,j)<\infty
$$
is valid. To the sequence of operators $\{\wh{A}_N\}$, there
corresponds a secondary quantized operator $\wh{A}$ in the Fock
space $\cH_B$.

The following assertion holds.

\hbox{}

{\bf Lemma}. {\it The symbol of the averaged operator
$\ov{\wh{A}}$ can be expressed as follows{\rm:}
\begin{eqnarray}
&&A(b_1^*(x_1^1,j^1),b_1(x_1^1,j^1),\dots,b_k^*(x_1^k,\dots,x_k^k,j^k),
b_k(x_1^k,\dots,x_k^k,j^k))=\nn\\
&&\qquad=
\Sp\left(\wh{A}\wh{\rho}\right)
\exp\Big(-\sum^{\infty}_{j=1}\int dz\,
b^*_1(z,j)b_1(z,j)-\nn\\
&&\qquad-\dots-\sum^{\infty}_{j=1}\int\dots\int dz_1\dots dz_k\,
b^*_k(z_1,\dots,z_k,j)b_k(z_1,\dots,z_k,j)\Big),
\label{f20}
\end{eqnarray}
where
$\wh{\rho}$ is an operator in $\cH_B$ and has the
form}
\begin{eqnarray*}
&&\wh{\rho}=\sum_{N_1=0}^{\infty}\dots\sum_{N_k=0}^{\infty}
\frac1{N_1!\dots N_k!}\prod_{l=1}^k\Bigl(\sum_{j^l=0}^{\infty}
\int\dots\int dx_1^ldy_1^l\dots dx_l^ldy_l^l\times\nn\\
&&\qquad\times b_l(x_1^l,\dots,x_l^l,j^l)\str{2}{\wh{\psi}^+}(x_1^l)
\dots\str{2}{\wh{\psi}^+}(x_l^l)\times\nn\\
&&\qquad\times b_l^*(y_1^l,\dots,y_l^l,j^l)\str{1}{\wh{\psi}^-}(y_1^l)
\dots\str{1}{\wh{\psi}^-}(y_l^l)\Bigr)^{N_l}
\exp\left(-\int dz\str{2}{\wh{\psi}^+}(z)
\str{1}{\wh{\psi}^-}(z)\right).
\end{eqnarray*}

\hbox{}

This lemma follows from a lemma proved in [3].

As was shown in the Introduction,
it is impossible to find how metastable states depend on the
temperature, if the quantization of free energy is not introduced.

In the simples case of microcanonical distribution,
we have the following expression for free energy $F$:
\begin{equation}\label{*}
F=\lambda-\theta\ln n_{\lambda}
\end{equation}
where $\theta$ is the temperature,
$\lambda$ is the energy level,
$n_{\lambda}$ is its multiplicity,
and $S=\ln n_{\lambda}$ is entropy by definition.
Let $d E_{\lambda}$ be the spectral family of
a self-adjoint energy operator $\wh{H}$ with discrete spectrum.
Then, as is known, we have
$$
\wh{H}=\int\,\lambda dE_{\lambda}.
$$
It is natural to define the quantum free energy as
\begin{equation}\label{**}
\wh{F}=\int\,(\lambda-\theta\ln n_{\lambda})dE_{\lambda}.
\end{equation}
As a rule,
in statistical physics we deal with the canonical distribution.
The passage from (5) to the canonical distribution is not
trivial and here can be carried out only in the simplest cases.
Similarly,
the method for obtaining the values of quantum
entropy, which we propose later,
can be derived from (6) not in the general case
in which we shall use it.
This separate problem lies beyond the framework of the present work.

Next, let $\ov{\wh{\cH}}$ and $\ov{\wh{E}}$
be the Hamiltonian and
the identity operator, respectively, averaged by this method.
By the lemma, the symbols of these operators, up to a factor,
are equal to $\Sp\wh{\rho}$ and $\Sp(\wh{H}\wh{\rho})$,
respectively. If $\wh{\rho}$ is
treated as a statistical operator (a density matrix), then
$\Sp\wh{\rho}$ is the norm of this operator and
$\Sp(\wh{H}\wh{\rho})$ is the mean value of the
Hamiltonian multiplied by $\Sp \widehat{\rho}$.
{}Furthermore,  entropy corresponding to $\wh{\rho}$ is given
by
$$
S=Sp\left(\wh{\rho}\ln\left(\frac{\wh{\rho}}{\Sp\wh{\rho}}\right)\right)
\left(\Sp\wh{\rho}\right)^{-1}.
$$
By analogy with statistical physics,
we refer to the functional
\begin{eqnarray}
&&S(b_1^*(x_1^1,j^1),b_1(x_1^1,j^1),\dots,b_k^*(x_1^k,\dots,x_k^k,j^k),
b_k(x_1^k,\dots,x_k^k,j^k))=\nn\\
&&\qquad=\Sp\left(\wh{\rho}\ln\left(\frac{\wh{\rho}}
{\Sp\wh{\rho}}\right)\right)\exp\Big(-\sum^{\infty}_{j=1}\int dz\,
b^*_1(z,j)b_1(z,j)-\dots-\nn\\
&&\qquad-\sum^{\infty}_{j=1}\int\dots\int dz_1\dots dz_k\,
b^*_k(z_1,\dots,z_k,j)b_k(z_1,\dots,z_k,j)\Big)
\end{eqnarray}
as the {\it symbol of averaged entropy}. To this symbol, we
assign the following operator in $\cF$:
\begin{equation}
\ov{\wh{S}}={\rm Reg}S(\str{2}{\wh{b}_1^+}(x_1^1,j^1),
\str{1}{\wh{b}_1^-}(x_1^1,j^1),\dots,\str{2}{\wh{b}_k^+}
(x_1^k,\dots,x_k^k,j^k),\str{1}{\wh{b}_k^-}(x_1^k,\dots,x_k^k,j^k)),
\end{equation}
where ${\rm Reg}$ stands for the regularization if necessary.

\hbox{}

{\bf Definition}.
Numbers $\lambda$ such that
\begin{equation}
(\ov{\wh{\cH}}+\theta\ov{\wh{S}})\Phi
= \lambda\ov{\wh{E}}\Phi,
\qquad\Phi\in\cF, \qquad \Phi \not \equiv 0,
\end{equation}
are called {\it quantum eigenvalues of free energy} for a
system of bosons with Hamiltonian $\wh{H}$ at temperature
$\theta$ for $k$ cluster types.

\section{Supersecondary quantization for fermions}

Now we introduce the averaging of operators
and the quantum free energy for fermions. We assume that the
fermions may have a spin variable
$s$ ranging in a discrete finite set $\Sigma$. Consider the
Hilbert space $\cF_{N_1,\dots,N_k}$ of functions (1),
where $x^l_{p_l}\in \bT^3\times\Sigma$.
Throughout this section, we use the notation
\begin{equation}
\int dx=\sum_{s\in\Sigma}\int d\xi,
\end{equation}
where $x=(\xi,s)$, $\xi\in\bT^3$, $s\in\Sigma$.
Let us define the fermion space $\cF^F_{N_1,\dots,N_k}$.

\hbox{}

{\bf Definition}.
The space $\cF^F_{N_1,\dots,N_k}$ is the subspace of
$\cF_{N_1,\dots,N_k}$ formed by elements
$\cF_{N_1,\dots,N_k}$ that are antisymmetric with respect to
the permutations of arbitrary
$x^l_p$ and $x^m_q$, $l,m=1,\dots,k$, 
$p=1,\dots,lN_l$, $q=1,\dots,mN_m$.
The projection on this subspace will be denoted by
$\wh{P}^F_{N_1,\dots,N_k}$:
$$
\wh{P}^F_{N_1,\dots,N_k}:\,\, \cF_{N_1,\dots,N_k}\to \cF^B_{N_1,\dots,N_k}.
$$

\hbox{}

Let us introduce the notion of an averaged operator in the
fermion case. Let $\{\wh{A}_N\}$ be a sequence of operators in
the subspaces
$L_{2} (\underbrace{ \bT^3\times \Sigma \times \dots \times
\bT^3\times\Sigma)}_{N \mbox{ times}}$ formed by elements
$\psi (x_1, \dots, x_N)$ antisymmetric with
respect to the permutations of the variables
$x_j$ and $x_k$, $x_j\in \bT^3\times\Sigma$.
To these operators, we assign the operators
$\wt{\wh{A}}_{N_1,\dots,N_k}$, $N_1+2N_2+\dots+kN_k=N$,
defined on the subspaces $\cF^F_{N_1,\dots,N_k}$
and acting as the identity operators
with respect to the discrete variables $j^l_p$.

\hbox{}

{\bf Definition}. The {\it averaged operator} $\ov{\wh{A}}$
is the operator
\begin{equation}
\ov{\wh{A}}
=\sum^{\infty}_{N_1=0}\dots\sum^{\infty}_{N_k=0}
(N_1+\dots+kN_k)!
\wh{i}_{N_1,\dots,N_k}
\wt{\wh{A}}_{N_1+\dots+kN_k}
\wh{P}^F_{N_1,\dots,N_k}\wh{P}_{N_1,\dots,N_k}
\end{equation}
in the Fock space $\cF$.

\hbox{}

{\bf Remark}.
The operator $\wh{i}_{N_1,\dots,N_k}
\wt{\wh{A}}_{N_1+\dots+kN_k}
\wh{P}^F_{N_1,\dots,N_k}\wh{P}_{N_1,\dots,N_k}$
can be written as
\begin{eqnarray}
&&
\frac1{N_1!\dots N_k!}
\sum^{\infty}_{j^1_1=1}\dots\sum^{\infty}_{j^1_{N_1}=1}
\dots\sum^{\infty}_{j^k_1=1}\dots\sum^{\infty}_{j^k_{N_k}=1}
\int\dots\int
dx^1_1\dots dx^1_{N_1}\dots dx^k_1\dots dx^k_{kN_k}
\nn\\
&&\qquad
\times
\stackrel{2}{\wh{b}^+_1}(x^1_1,j^1_1)
\dots
\stackrel{2}{\wh{b}^+_1}(x^1_{N_1},j^1_{N_1})
\dots
\stackrel{2}{\wh{b}^+_k}(x^k_1,\dots,x^k_k,j^k_k)
\dots
\nn\\
&&\qquad
\times
\stackrel{2}{\wh{b}^+_k}(x^k_{kN_k-k+1},\dots,x^k_{kN_k},j^k_{N_k})
\wh{A}_{N_1+\dots+kN_k}
\nn\\
&&\qquad
\times
\Asymm_{x^1_1\dots x^1_{N_1}\dots x^k_1\dots x^k_{kN_k}}
\Big\{
\stackrel{1}{\wh{b}^-_1}(x^1_1,j^1_1)
\dots
\stackrel{1}{\wh{b}^-_1}(x^1_{N_1},j^1_{N_1})
\dots
\nn\\
&&\qquad\qquad\qquad\qquad\qquad
\times
\stackrel{1}{\wh{b}^-_k}(x^k_1,\dots,x^k_k,j^k_k)
\dots
\stackrel{1}{\wh{b}^-_k}(x^k_{kN_k-k+1},\dots,x^k_{kN_k},j^k_{N_k})
\Big\}
\nn\\
&&\qquad
\times
\exp\Big(
-\sum^{\infty}_{j=1}\int dz\,
\stackrel{2}{\wh{b}^+_1}(z,j)\stackrel{1}{\wh{b}^-_1}(z,j)
-\dots
\nn\\
&&\qquad\qquad\qquad
-\sum^{\infty}_{j=1}\int\dots\int dz_1\dots dz_k\,
\stackrel{2}{\wh{b}^+_k}(z_1,\dots,z_k,j)
\stackrel{1}{\wh{b}^-_k}(z_1,\dots,z_k,j)
\Big),
\end{eqnarray}
where
$\Asymm_{x_1\dots x_N}$ is the operator of
antisymmetrization with respect to the variables
$x_1,\dots,x_N$. This expression is the fermion counterpart of
(2).

\hbox{}

The {\it symbol} of the averaged operator $\ov{\wh{A}}$ in the
fermion case is defined as the functional
\begin{eqnarray}
&&A(b_1^*(x_1^1,j^1),b_1(x_1^1,j^1),\dots,b_k^*(x_1^k,\dots,x_k^k,j^k),
b_k(x_1^k,\dots,x_k^k,j^k))=\nn\\
&&\qquad=\sum_{N_1=0}^{\infty}\dots\sum_{N_k=0}^{\infty}
\frac{(N_1+2N_2+\dots+kN_k)!}{N_1!N_2!\dots N_k!}\times\nn\\
&&\qquad\times\sum^{\infty}_{j^1_1=1}\dots\sum^{\infty}_{j^1_{N_1}=1}
\dots\sum^{\infty}_{j^k_1=1}\dots\sum^{\infty}_{j^k_{N_k}=1}
\int\dots\int
dx^1_1\dots dx^1_{N_1}\dots dx^k_1\dots dx^k_{kN_k}\times\nn\\
&&\qquad\times b^*_1(x^1_1,j^1_1)\dots b^*_1(x^1_{N_1},j^1_{N_1})
\dots b^*_k(x^k_1,\dots,x^k_k,j^k_k)\dots
b^*_k(x^k_{kN_k-k+1},\dots,x^k_{kN_k},j^k_{N_k})\times\nn\\
&&\qquad\times\wh{A}_{N_1+\dots+kN_k}
\Asymm_{x^1_1\dots x^1_{N_1}\dots x^k_1\dots x^k_{kN_k}}
\Big\{ b_1(x^1_1,j^1_1)\dots b_1(x^1_{N_1},j^1_{N_1})\dots\times\nn\\
&&\qquad\times b_k(x^k_1,\dots,x^k_k,j^k_k)\dots
b_k(x^k_{kN_k-k+1},\dots,x^k_{kN_k},j^k_{N_k})\Big\}\nn\\
&&\qquad\times\exp\Big(
-\sum^{\infty}_{j=1}\int dz\,
b^*_1(z,j)b_1(z,j)-\dots-\nn\\
&&\qquad-\sum^{\infty}_{j=1}\int\dots\int dz_1\dots dz_k\,
b^*_k(z_1,\dots,z_k,j)b_k(z_1,\dots,z_k,j)\Big),
\end{eqnarray}
where
$b_l(\cdot,\dots,\cdot,j^l)\in L^2(\bT^{3l})$, $l=1,\dots,k$,
$j^l=0,1,\dots$, and the inequality
$$
\sum_{j=0}^{\infty}\int\dots\int dx_1\dots dx_l
b_l^*(x_1,\dots,x_l,j)
b_l(x_1,\dots,x_l,j)<\infty
$$
is valid. To the sequence $\{\wh{A}_N\}$ of operators, there
corresponds a secondary quantized operator $\wh{A}$ in the
fermion Fock space $\cH_F$, which consists of sequences of
antisymmetric functions and is an analog of the space $\cH_B$
(see [4] for details).

The following assertion holds for the fermion case.

\hbox{}

{\bf Lemma}. {\it The symbol of the averaged operator
$\ov{\wh{A}}$ is given by the formula
\begin{eqnarray}
&&A(b_1^*(x_1^1,j^1),b_1(x_1^1,j^1),\dots,b_k^*(x_1^k,\dots,x_k^k,j^k),
b_k(x_1^k,\dots,x_k^k,j^k))=\nn\\
&&\qquad=
\Sp\left(\wh{A}\wh{\rho}\right)
\exp\Big(-\sum^{\infty}_{j=1}\int dz\,
b^*_1(z,j)b_1(z,j)-\nn\\
&&\qquad-\dots-\sum^{\infty}_{j=1}\int\dots\int dz_1\dots dz_k\,
b^*_k(z_1,\dots,z_k,j)b_k(z_1,\dots,z_k,j)\Big),
\end{eqnarray}
where $\wh{\rho}$ is the operator in $\cH_F$ given by
\begin{eqnarray}
&&\wh{\rho}=\sum_{N_1=0}^{\infty}\dots\sum_{N_k=0}^{\infty}
\frac1{N_1!\dots N_k!}\sum_{j^1_1=0}^{\infty}
\dots\sum_{j^1_{N_1}=0}^{\infty}
\dots\sum_{j^k_1=0}^{\infty}\dots\sum_{j^k_{N_k}=0}^{\infty}\times\nn\\
&&\qquad\times\int\dots\int dx^1_1dy^1_1\dots dx^1_{N_1}dy^1_{N_1}\dots
dx^k_1dy^k_1\dots dx^k_{kN_k}dy^k_{kN_k}\wh{\psi}^+(x^1_1)\dots
\wh{\psi}^+(x^1_{N_1})\times\nn\\
&&\qquad\times\dots\wh{\psi}^+(x^k_1)\dots
\wh{\psi}^+(x^k_{kN_k})\wh{P}_0\times\nn\\
&&\qquad\times b_1(x^1_1,j^1_1)\dots b_1(x^1_{N_1},j^1_{N_1})\dots
b_k(x^k_1,\dots,x^k_k,j^k_1)\dots
b_k(x^k_{kN_k-k+1},\dots,x^k_{kN_k},j^k_{N_k})\times\nn\\
&&\qquad\times b_1^*(y^1_1,j^1_1)\dots b_1^*(y^1_{N_1},j^1_{N_1})\dots
b_k^*(y^k_1,\dots,y^k_k,j^k_1)\dots
b_k^*(y^k_{kN_k-k+1},\dots,y^k_{kN_k},j^k_{N_k})\times\nn\\
&&\qquad\times\wh{\psi}^-(y^k_{kN_k})\dots
\wh{\psi}^-(y^k_1)\dots\wh{\psi}^-(y^1_{N_1})\dots
\wh{\psi}^-(x^1_1);
\end{eqnarray}
here $\wh{P}_0$ is the projection on the vacuum vector in $\cH_F$.}

\hbox{}

In the fermion case, we define the {\it symbol of averaged
entropy} as the functional
\begin{eqnarray}
&&S(b_1^*(x_1^1,j^1),b_1(x_1^1,j^1),\dots,b_k^*(x_1^k,\dots,x_k^k,j^k),
b_k(x_1^k,\dots,x_k^k,j^k))=\nn\\
&&\qquad=Sp\left(\wh{\rho}\ln\left(\frac{\wh{\rho}}
{\Sp\wh{\rho}}\right)\right)\exp\Big(-\sum^{\infty}_{j=1}\int dz\,
b^*_1(z,j)b_1(z,j)-\dots-\nn\\
&&\qquad-\sum^{\infty}_{j=1}\int\dots\int dz_1\dots dz_k\,
b^*_k(z_1,\dots,z_k,j)b_k(z_1,\dots,z_k,j)\Big)
\end{eqnarray}
To this symbol, we assign the operator
\begin{equation}
\ov{\wh{S}}={\rm Reg}S(\str{2}{\wh{b}_1^+}(x_1^1,j^1),
\str{1}{\wh{b}_1^-}(x_1^1,j^1),\dots,\str{2}{\wh{b}_k^+}
(x_1^k,\dots,x_k^k,j^k),\str{1}{\wh{b}_k^-}(x_1^k,\dots,x_k^k,j^k))
\end{equation}
in the space $\cF$.

\hbox{}

{\bf Definition}
The numbers $\lambda$ such that
\begin{equation}
(\ov{\wh{\cH}}+\theta\ov{\wh{S}})\Phi
= \lambda\ov{\wh{E}}\Phi,
\qquad\Phi\in\cF, \qquad \Phi \not \equiv 0,
\end{equation}
are called {\it quantum eigenvalues of free energy} for a
system of fermions with Hamiltonian $\wh{H}$ at temperature
$\theta$ for $k$ cluster types.

\section{Asymptotic series as $N\to\infty$ for a fermion system}

{}Further, we assume that the secondary quantized fermion
Hamiltonian in the space~$\cH_F$ has the form
\begin{eqnarray}\label{20}
\wh{H}&=&\int\!\!\int dxdy\,\wh{\psi}^+(x)T(x,y)\wh{\psi}^-(y)\nn\\
&&+\frac12
\int\!\!\int\!\!\int\!\!\int dxdx'dydy'\,
\wh{\psi}^+(x)\wh{\psi}^+(x')V(x,x',y,y')\wh{\psi}^-(y')\wh{\psi}^-(y),
\end{eqnarray}
where, just as everywhere,
$x,x',y,y'=(\zeta,s)$, $\zeta\in\bT^3$, $s\in\Sigma$,
and the functions $T(x,y)$, $V(x,x',y,y')$ have the
properties
$$
T(x,y)=T^*(y,x),\qquad V(x,x',y,y')=V^*(y,y',x,x').
$$
Let us consider the supersecondary quantized equations for 
free energy of fermions in another representation.
Suppose that $k$ is the number of clusters of the first type,
$M$ is the number of clusters of the second type,
and there are no clusters of any other type.
We choose some functions
$\Psi(x,y)\in L_2((\bT^3\times\Sigma)\times(\bT^3\times\Sigma))$,
$\phi(x,j)\in L_2(\bT^3\times\Sigma)\times l_2$ such that
\begin{eqnarray}
&&\Psi(x,y)=-\Psi(y,x),\nn\\
&&k=\Sp\Biggl(\wh{G}^t-1+\Bigl(1+(1-\wh{G}^t)^{-1}\wh{R}
(1-\wh{G})^{-1}\wh{R}^+\Bigr)^{-1}\Biggr),\nn\\
&&M=\frac12\Sp\Biggl(\Bigl(1+(1-\wh{G}^t)^{-1}\wh{R}
(1-\wh{G})^{-1}\wh{R}^+\Bigr)^{-1}(1-\wh{G}^t)^{-1}\wh{R}
(1-\wh{G})^{-1}\wh{R}^+\Biggr),\label{21}
\end{eqnarray}
where $\wh{G}$, $\wh{R}$ are operators in $L_2(\bT^3\times\Sigma)$
of the form
\begin{eqnarray}
&&\wh{G}=(1+\wh{A}^t)^{-1}\Biggl(\wh{A}^t+\wh{B}^+(1+\wh{A})^{-1}\wh{B}
(1+\wh{A}^t)^{-1}
\Bigl(1+\wh{B}^+(1+\wh{A})^{-1}\wh{B}(1+\wh{A}^t)^{-1}\Bigr)^{-1}
\Biggr)\nn\\
&&\wh{R}=-(1+\wh{A})^{-1}\wh{B}(1+\wh{A}^t)^{-1}
\Bigl(1+\wh{B}^+(1+\wh{A})^{-1}\wh{B}(1+\wh{A}^t)^{-1}\Bigr)^{-1},
\label{22}
\end{eqnarray}
and $\wh{A}$, $\wh{B}$ are the following operators in
$L_2(\bT^3\times\Sigma)$:
$$
(\wh{A}u)(x)=\sum_{j=1}^{\infty}\int dy\,\vp(x,j)\vp^*(y,j)u(y),
\quad (\wh{B}u)(x)=\int dy\,\Psi(x,y)u(y),
$$
where the superscript $+$ stands for Hermitian conjugation
and~$t$ for transposition.
We choose an arbitrary complete orthonormal system of functions
$\Psi_a(x,y)$, $a=1,2,\dots$, in a subspace of the space
$L_2((\bT^3\times\Sigma)\times(\bT^3\times\Sigma))$
whose elements are functions orthogonal to~$\Psi(x,y)$,
and an arbitrary complete orthonormal system of functions
$\vp_c(x,j)$, $c=1,2,\dots$
in the subspace of the space
$L_2(\bT^3\times\Sigma)\times l_2$
consisting of functions orthogonal to $\vp(x,j)$.
The following vector system corresponding to these systems
is complete and orthonormal in the subspace $\cF_{k,M}$
of the space~$\cF$
\begin{eqnarray}
&&\Phi_{\{n\},\{m\}}=\prod_{a=1}^{\infty}\frac{1}{\sqrt{n_a!}}
\left(\int\!\!\int dx_1dx_2\, \Psi_a(x_1,x_2)
\wh{b}^+_2(x_1,x_2)
\right)^{n_a}\nn\\
&&\qquad\times
\prod_{c=1}^{\infty}\frac{1}{\sqrt{m_c!}}\left(
\sum_{j=1}^{\infty}\int dy\,\vp_c(y,j)\wh{b}^+_1(y,j)
\right)^{m_c}\times\nn\\
&&\qquad\times\frac1{C^{\wt{N}_2}\sqrt{\wt{N}_2!}}\left(
\int\!\!\int dxdx'\Psi(x,x')\wh{b}^+_2(x,x')\right)^{\wt{N}_2}\nn\\
&&\qquad\times
\frac1{D^{\wt{N}_1}\sqrt{\wt{N}_1!}}\left(\sum_{j'=1}^{\infty}
\int dy'\vp(y',j')\wh{b}^+_1(y',j')\right)^{\wt{N}_1}\Phi_0,
\label{23}
\end{eqnarray}
where $\Phi_0$ is the vacuum vector of the space~$\cF$;
$\{n\}$, $\{m\}$ are sets of integers
$n_a\geq0$, $a=1,2,\dots$, and $m_c\geq0$, $c=1,2,\dots$,
such that
$$
\sum_{a=1}^{\infty}n_a\leq M,\qquad
\sum_{c=1}^{\infty}m_c\leq k,
$$
and the following notation is used:
$$
\wt{N}_2=M-\sum_{a=1}^{\infty}n_a,\qquad
\wt{N}_1=k-\sum_{c=1}^{\infty}m_c,
$$
and
$$
C=\sqrt{\int\!\!\int dxdx'\,|\Psi(x,x')|^2},\qquad
D=\sqrt{\sum_{j=1}^{\infty}\int dx\,|\vp(x,j)|^2}.
$$
We introduce the Fock space $\cM$ generated by the vacuum
vector~$Y_0$ and the following boson creation and annihilation
operators of two kinds:
$\wh{\beta}_a^{\pm}$, $a=1,2,\dots$, $\wh{d}_c^{\pm}$, $c=1,2,\dots$.
To each vector~(22) we assign the following vector of the
space~$\cM$:
\begin{equation}\label{24}
Y_{\{n\},\{m\}}=\prod_{a=1}^{\infty}\frac1{\sqrt{n_a!}}(\wh{\beta}_a^+)^{n_a}
\prod_{c=1}^{\infty}\frac1{\sqrt{m_c!}}(\wh{d}_c^+)^{m_c}Y_0.
\end{equation}
Taking into account the relation between (22) and (23),
we can write the equation for quantum values of  free energy
of fermions in the space~$\cM$ as
\begin{equation}\label{25}
\wh{F}_{\theta}Y=\lambda\wh{E}Y,\qquad Y\in\cM,\qquad Y\ne0,
\end{equation}
where $\wh{F}_{\theta}$, $\wh{E}$ are the following operators in $\cM$:
\begin{eqnarray*}
&&\wh{F}_{\theta}=\exp\left(-\sum_{l=1}^{\infty}\str{2}{\wh{\beta}_l^+}
\str{1}{\wh{\beta}_l^-}-\sum_{p=1}^{\infty}\str{2}{\wh{d}_p^+}
\str{1}{\wh{d}_p^-}\right)\sum_{\{n\},\{m\}}\sum_{\{n'\},\{m'\}}
\left(\Phi_{\{n\},\{m\}},(\ov{\wh{\cH}}+\theta\ov{\wh{S}})
\Phi_{\{n'\},\{m'\}}\right)\times\\
&&\qquad\times\prod_{a=1}^{\infty}\frac
{(\str{2}{\wh{\beta}_a^+})^{n_a}(\str{1}{\wh{\beta}_a^-})^{n_a'})}
{\sqrt{n_a!n_a'!}}
\prod_{c=1}^{\infty}\frac
{(\str{2}{\wh{d}_c^+})^{m_c}(\str{1}{\wh{d}_c^-})^{m_c'})}
{\sqrt{m_c!m_c'!}},\\
&&\wh{E}=\exp\left(-\sum_{l=1}^{\infty}\str{2}{\wh{\beta}_l^+}
\str{1}{\wh{\beta}_l^-}-\sum_{p=1}^{\infty}\str{2}{\wh{d}_p^+}
\str{1}{\wh{d}_p^-}\right)\sum_{\{n\},\{m\}}\sum_{\{n'\},\{m'\}}
\left(\Phi_{\{n\},\{m\}},\ov{\wh{E}}\Phi_{\{n'\},\{m'\}}\right)\times\\
&&\qquad\times\prod_{a=1}^{\infty}\frac
{(\str{2}{\wh{\beta}_a^+})^{n_a}(\str{1}{\wh{\beta}_a^-})^{n_a'})}
{\sqrt{n_a!n_a'!}}
\prod_{c=1}^{\infty}\frac
{(\str{2}{\wh{d}_c^+})^{m_c}(\str{1}{\wh{d}_c^-})^{m_c'})}
{\sqrt{m_c!m_c'!}}.
\end{eqnarray*}
We assume that equation (24) has the solutions
\begin{equation}\label{26}
Y=\sum_{\{n\},\{m\}}\gamma_{\{n\},\{m\}}Y_{\{n\},\{m\}},
\end{equation}
such that in the limit as $N\to\infty$, $L\to\infty$,
$N/L^3\to{\rm const}$
we have
$$
\frac{\sum_{\{n\},\{m\}}(\sum_{a=1}^{\infty}n_a)|\gamma_{\{n\},\{m\}}|^2}
{\sum_{\{n\},\{m\}}|\gamma_{\{n\},\{m\}}|^2}=O(1),\qquad
\frac{\sum_{\{n\},\{m\}}(\sum_{c=1}^{\infty}m_c)|\gamma_{\{n\},\{m\}}|^2}
{\sum_{\{n\},\{m\}}|\gamma_{\{n\},\{m\}}|^2}=O(1).
$$
To such solutions of equation (24)
there correspond the quantum value of free energy
\begin{eqnarray}
&&\lambda(\Psi(x,y),\vp(x,j))=\int\!\!\int dxdyT(x,y)G(x,y)+\nn\\
&&\qquad +
\frac12\int\!\!\int\!\!\int\!\!\int dxdx'dydy'V(x,x',y,y')
\Bigl(G(x,y)G(x',y')-\nn\\
&&\qquad-G(x,y')G(x',y)+R^*(x',x)R(y',y)\Bigr)+\nn\\
&&\qquad+\theta\Sp f
\left(
\begin{array}{cc}
\wh{G}-\frac12&\wh{R}^+\\
\wh{R}&-\wh{G}^t+\frac12
\end{array}
\right)+O(1),
\label{27}
\end{eqnarray}
where $f(\xi)$, $\xi\in\bR$ has the form
$$
f(\xi)=\frac12(\frac12+\xi)\ln(\frac12+\xi)+
\frac12(\frac12-\xi)\ln(\frac12-\xi),
$$
and the functions $G(x,y)$, $R(x,y)$ are the kernels of the
operators (21):
$$
(\wh{G}u)(x)=\int dy\,G(x,y)u(y),\quad
(\wh{R}u)(x)=\int dy\,R(x,y)u(y).
$$
The vector $\Phi\in\cF$ corresponding to (\ref{26})
we denote by $\Phi(\Psi(x,y),\varphi(x,j))$.
{}For two distinct pairs of functions
$\Psi(x,y)$, $\varphi(x,j)$ and $\Psi'(x,y)$, $\varphi'(x,j)$
satisfying conditions (\ref{21}),
the following matrix element of an arbitrary
supersecondary quantized operator $\ov{\wh{A}}$,
which is bounded in~$N$:
$$
\frac{\left(\Phi(\Psi'(x,y),\vp'(x,j)),
\ov{\wh{A}}\Phi(\Psi(x,y),\vp(x,j))\right)}
{\left(\Phi(\Psi'(x,y),\vp'(x,j)),\Phi(\Psi(x,y),\vp(x,j))\right)}
$$
is exponentially small in~$N$.

Therefore, in the case in which
the values $\lambda(\Psi(x,y),\varphi(x,j))$ and
$\lambda(\Psi'(x,y),\varphi'(x,j))$
corresponding to these function pairs
do not coincide,
the vectors $\Phi(\Psi(x,y),\varphi(x,j))$ and
$\Phi(\Psi'(x,y),\varphi'(x,j))$ belong to different series.
If the resonance occurs, i.e., the value (\ref{27}) is the same
for $\Psi(x,y)$, $\varphi(x,j)$ and $\Psi'(x,y)$, $\varphi'(x,j)$,
then the series is formed by linear combinations of the vectors
$\Phi(\Psi(x,y),\varphi(x,j))$ and $\Phi(\Psi'(x,y),\varphi'(x,j))$.
{}From all series of solutions of the equation for quantum values
of free energy corresponding to some chosen~$k$ and~$M$,
we choose the series with the least value (\ref{27}).
The functions $\Psi(x,y)$, $\varphi(x,j)$ corresponding to this
series are determined from the equation of self-consistent type:
\begin{equation}\label{28}
\left(
\begin{array}{cc}
\wh{G}-\frac12&\wh{R}^+\\
\wh{R}&-\wh{G}^t+\frac12
\end{array}
\right)=
-\frac12{\rm th}\left(\frac1{2\theta}\left(
\begin{array}{cc}
\wh{\wt{T}}^t&\wh{\wt{V}}^+\\
\wh{\wt{V}}&-\wh{\wt{T}}
\end{array}
\right)\right)
\end{equation}
where the operators $\wh{\wt{T}}$, $\wh{\wh{V}}$ are expressed in
terms of $\wh{G}$, $\wh{R}$
as follows:
\begin{eqnarray}\label{29}
&&\wh{\wt{T}}=\wh{X}-\mu+(2\mu-\omega)(1-\wh{G}^t)^{-1}\wh{R}
(1-\wh{G})^{-1}\wh{R}^+(1-\wh{G}^t)^{-1}\nn\\
&&\qquad \times
\Bigl(1+\wh{R}(1-\wh{G})^{-1}\wh{R}^+(1-\wh{G}^t)^{-1}\Bigr)^{-2},\nn\\
&&\wh{\wt{V}}=\wh{Z}+(2\mu-\omega)(1-\wh{G}^t)^{-1}\wh{R}
(1-\wh{G})^{-1}\Bigl(1+\wh{R}^+
(1-\wh{G}^t)^{-1}\wh{R}(1-\wh{G})^{-1}\Bigr)^{-2},
\end{eqnarray}
and the operators $\wh{X}$, $\wh{Z}$
in the space $L_2(\bT^3\times l_2)$ have
the form
\begin{eqnarray}\label{30}
&&(\wh{X}u)(x)=\int dy\,T(x,y)u(y)+\int\!\!\!\int\!\!\!\int dx'dydy'
(V(x,x',y,y')-V(x,x',y',y))G(x',y')u(y),\nn\\
&&(\wh{Z}u)(x)=\int\!\!\int\!\!\int dx'dydy'V(x',x,y,y')R(y',y)u(x').
\end{eqnarray}
The numbers $\mu$, $\omega$ in (\ref{29}) are determined by
conditions (\ref{21}).
The minimal quantum value of free energy (\ref{27})
for some chosen~$k$ and~$M$ we denote by $\lambda_{k,M}$.
Let us consider the minimum of~$\lambda_{k,M}$
with respect to~$k$ and~$M$ provided $k+2M=N={\rm const)}$.
We denote this minimum by~$\lambda_N$.
The functions $\Psi(x,y)$, $\varphi(x,j)$
corresponding to this minimum can be found from the equation
\begin{equation}\label{31}
\left(
\begin{array}{cc}
\wh{G}-\frac12&\wh{R}^+\\
\wh{R}&-\wh{G}^t+\frac12
\end{array}
\right)=
-\frac12{\rm th}\left(\frac1{2\theta}\left(
\begin{array}{cc}
\wh{X}^t-\mu&\wh{Z}^+\\
\wh{Z}&-\wh{X}+\mu
\end{array}
\right)\right)
\end{equation}
where the operators $\wh{X}$, $\wh{Z}$ are the same as (\ref{30})
and the number~$\mu$ is determined by the condition $k+2M=N$,
where~$k$ and~$M$ have the form (\ref{21}).
Equation (\ref{31}) is well known.
It coincides with the Bardeen--Cooper--Schriffer--Bogolyubov
equation in the theory of superconductivity.
The solutions of this equation with different~$\theta$
determine the functions $k(\theta)$, $M(\theta)$
by formulas (\ref{21})
and the function $\lambda_N(\theta)$ by formula (\ref{27}).
The temperature~$\theta_c$ at which $M(\theta_c)=0$
is called critical.
At this temperature the heat capacity
\begin{equation}\label{32}
C=-\theta\lambda_N''(\theta)
\end{equation}
has a discontinuity.
Since resonances can occur for $\lambda_{k,M}=\lambda_{k',M'}$
with $k\ne k'$ and $M\ne M'$,
we can replace the functions $k(\theta)$, $M(\theta)$
by some other $k_1(\theta)$, $M_1(\theta)$
such that $M_1(\theta_1)=0$ for $\theta_1>\theta_c$
and thus to increase the critical temperature.
To calculate the heat capacity corresponding to a temperature
series distinct from the Bardeen--Cooper--Schriffer--Bogolyubov-series,
we need to replace $\lambda_N(\theta)$
by $\lambda_{k(\theta),M(\theta)}$ in formula (\ref{32}).

\section{Asymptotic series as $N\to\infty$ for a boson series}

In a similar way, we can find
series of quantum values of free energy for bosons
in the case in which the number of cluster types is $2$.

Here we present an asymptotics for the case
in which the condensate can exist.

The principal state of the series is expressed in the same way
as for fermions via the vectors of some system of the form (21).
{}For the sets $\{n\}$, $\{m\}$ the same condition as that for
fermions holds in the thermodynamical limit.
Let us study the series such that
\begin{equation}
\Psi(x,y)=\frac{1}{L^3}\sum_p\phi(p)e^{ip(x-y)}, \quad\phi(p)=\phi(-p),
\qquad\vp(x,j)=\sqrt{\frac{a(p_j)}{L^3}}e^{ip_jx},
\end{equation}
where $p$ are three-dimensional vectors of the form
$2\pi(l_1,l_2,l_3)/L$, $l_1,l_2,l_3\in\bZ$,
$\sum_{p}$ stands for the summation over all such vectors,
and $p_j$ is a single-valued mapping of the set
$j=1,2,\dots$ on the set of such vectors.
Moreover,
$\phi(q)$, $a(q)$ are continuous functions of the variable
$q\in\bR^3$ such that $\phi(q)=\phi(-q)$,
$a(q)=a^*(q)$, $0\leq a(q)\leq 1$.
If the condensate consists of particles with zero momentum,
then the functions $\phi(q)$, $a(q)$ satisfy the inequality
\begin{equation}
(1-a(q))(1-a(-q))-\phi^*(q)\phi(q)\geq0.
\end{equation}
Note that we have the equality only for $q=0$.
In this case the asymptotics of the ratio of the quantum value
of free energy to the volume in the thermodynamical limit has
the form
\begin{eqnarray}
&&\frac{\lambda}{L^3}=\frac1{(2\pi)^3}\int dq\,\frac{\hb^2q^2}{2m}G(q)+
\frac{U_0a^3n_0\wt{v}(0)}{2}+\frac{U_0a^3n_0}{(2\pi)^3}\int dq\,(\wt{v}(0)+
\wt{v}(aq))G(q)+\nn\\
&&+\frac{U_0a^3n_0}{2(2\pi)^3}\int dq\,\wt{v}(aq)(R(q)+R^*(q))+
\frac{U_0a^3}{2(2\pi)^6}\int\!\!\int dqdl\,(\wt{v}(0)+\wt{v}(a(q-l)))
G(q)G(l)+\nn\\
&&\qquad+\frac{U_0a^3}{2(2\pi)^6}\int\!\!\int dqdl\,\wt{v}(a(q-l))
R^*(q)R(l)+\nn\\
&&\qquad +\frac{\theta}{(2\pi)^3}\int dq\,\Bigl(n(q)\ln n(q)-
(1+n(q))\ln(1+n(q))\Bigr),
\end{eqnarray}
where $n_0$ is the condensate density,
$$
n_0=\frac{N}{L^3}-\frac1{(2\pi)^3}\int dq\,G(q),
$$
$N$ is the number of particles, and just as for fermions we have
$N=k+2M$. The functions
$G(q)$, $R(q)$, $n(q)$ in formula (34) are expressed via
$a(q)$, $\phi(q)$ as follows
\begin{eqnarray}
&&G(q)=\frac{a(q)(1-a(-q))+\phi^*(q)\phi(q)}
{(1-a(q))(1-a(-q))-\phi^*(q)\phi(q)},\nn\\
&&R(q)=\frac{\phi(q)}{(1-a(q))(1-a(-q))-\phi^*(q)\phi(q)},\nn\\
&&n(q)=\frac12\sqrt{(1+G(q)+G(-q))^2-4R^*(q)R(q)}-\frac12(1-G(q)+G(-q)).
\end{eqnarray}
Expression (34) corresponds to the quantum free energy
of the principal state of some series if the functions $\phi(q)$,
$a(q)$ satisfy the relations
\begin{eqnarray}
&&
k=L^3n_0a(0)+\frac{L^3}{2(2\pi)^3}\int dq\,\frac{a(q)+a(-q)-2a(q)a(-q)}
{(1-a(q))(1-a(-q))-\phi^*(q)\phi(q)},\nn\\
&&
M=\frac{L^3n_0}{2}(1-a(0))+\frac{L^3}{2(2\pi)^3}\int dq\,
\frac{\phi^*(q)\phi(q)}{(1-a(q))(1-a(-q))-\phi^*(q)\phi(q)}
\end{eqnarray}
and (34) attains the minimum on these functions
provided that conditions (36) are satisfied.
One can easily see that if $n_0\ne0$ then
$$
\left.\frac{\delta}{\delta\phi^*(q)}\frac{\lambda}{L^3}\right|_{\phi=0}=
\frac{U_0a^3n_0}{2(2\pi)^3}\frac{\wt{v}(aq)}{(1-a(q))(1-a(-q))}\ne0.
$$
This implies that for any function $a(p)$ there always exists
a function $\phi(p)$ such that the value of free energy
(34) determined by the functions $a(p)$, $\phi(p)$ is less than
the value of free energy for $a(p)$, $\phi(p)=0$.
Although the nature can ``set'' the system on any metastable
state, the physicists always assume that the state with the
least energy is realized.
Hence here we do not consider boson systems
with clusters only of the first type and we always assume that
there exists at least two types of clusters.

\section{Variational method for the supersecondary quantized
problem}

We consider a variational method which allows us
to construct successive approximates to the solutions
of the supersecondary quantized equations for the quantum free
energy both in the boson and fermion cases.
We study the fermion case in detail.

Suppose that the number of clusters of the first type
is $k$, the number of clusters of the second type is $M$,
and there are no clusters of any other type.
We study some complete system of functions $\Psi_a(x,y)$,
$a=0,1,\dots$, in the space $L_2((\bT^3\times\Sigma)
\times(\bT^3\times\Sigma))$ and some complete system of functions
$\vp_c(x,j)$, $c=0,1,\dots$, in the space
$L_2(\bT^3\times\Sigma)\times l_2$.
In this section we choose function systems no in the same way as
in Sect.~3. Below we explain how we choose these function
systems.

To these functions, by formula (22) we assign
a vector system that is complete in the subspace $\cF_{k,M}$
of the space $\cF$.
We assume that the numbers $k$ and $M$ are sufficiently large
so that it is possible to regulate the entropy operator
$\ov{\wh{S}}$. Let us consider the sequence
\begin{equation}\label{42}
\lambda_b=\frac{\left(\wt{\Phi}_b,\left(\ov{\wh{\cH}}+\theta\ov{\wh{S}}\right)
\wt{\Phi}_b\right)}{\left(\wt{\Phi}_b,\ov{\wh{E}}\wt{\Phi}_b\right)},
\quad b=0,1,\dots,
\end{equation}
where $\wt{\Phi}_b\in\cF$ is a vector of the form (22)
corresponding to the set of numbers
$n_a=M\delta_{ab}$, $m_c=k\delta_{cb}$,
and $\delta_{ab}$ is the Kronecker delta.
We choose function systems $\Psi_a(x,y)$, $a=0,1,\dots$, and
$\vp_c(x,j)$, $c=0,1,\dots$, so that the following conditions hold:
\begin{equation}\label{43}
\frac{\delta\lambda_b}{\delta\Psi_b(x,y)}=0,\qquad
\frac{\delta\lambda_b}{\delta\vp_b(x,j)}=0
\end{equation}
and, in addition,
$$
\min_{b}\lambda_b=\lambda_0.
$$
Conditions (\ref{43}) are
Hartree-Fock--Bogolyubov--Bardeen--Cooper--Schriffer type
equations for the functions $\Psi_b(x,y)$, $\vp_b(x,j)=0$.

{}For example, for $\Psi(x,y)=0$ at zero temperature, 
we can write the equation for the function $\vp_b(x,j)$ 
as a system of Hartree--Fock type equations for the functions 
$u_j(x)=\vp_b(x,j)$.

In the equation for the quantum free energy
we expand the vector $\Phi$ with respect to the vector system (22):
\begin{equation}\label{44}
\Phi=\sum_{\{n\},\{m\}}A_{\{n\}\{m\}}\Phi_{\{n\},\{m\}},
\end{equation}
where
$$
\sum_{\{n\},\{m\}}
$$
stands for the summation over all possible sets of $\{n\}$, $\{m\}$.
Taking into account (\ref{44}), we can rewrite the equation for free
energy as the linear system
\begin{equation}\label{45}
\sum_{\{n\},\{m\}}F_{\{n'\}\{m'\},\{n\}\{m\}}A_{\{n\}\{m\}}=
\lambda\sum_{\{n\},\{m\}}E_{\{n'\}\{m'\},\{n\}\{m\}}A_{\{n\}\{m\}},
\end{equation}
where the coefficients
$F_{\{n'\}\{m'\},\{n\}\{m\}}$, $E_{\{n'\}\{m'\},\{n\}\{m\}}$
are expressed via the supersecondary quantized operators
$\ov{\wh{\cH}}$, $\ov{\wh{S}}$, $\ov{\wh{E}}$ and the vectors (22).
Equation (\ref{45}) is always the exact representation for the
equation of quantum free energy.
Approximate solutions of the equation for free energy can be
obtained from the linear system
\begin{equation}\label{46}
\sum_{\{n\},\{m\}}^{\wt{k},\wt{M}}F_{\{n'\}\{m'\},\{n\}\{m\}}A_{\{n\}\{m\}}=
\lambda\sum_{\{n\},\{m\}}^{\wt{k},\wt{M}}E_{\{n'\}\{m'\},\{n\}\{m\}}
A_{\{n\}\{m\}},
\end{equation}
where $0\leq\wt{k}\leq k$, $0\leq\wt{M}\leq M$ are integers and
$$
\sum_{\{n\},\{m\}}^{\wt{k},\wt{M}}
$$
stands for summation over the sets of numbers
$\{n\}$, $\{m\}$ such that
\begin{equation}
\sum_{a=1}^{\infty}n_a\leq\wt{M},\qquad
\sum_{c=1}^{\infty}m_c\leq\wt{k}.
\end{equation}
{}For $\wt{k}=0$, $\wt{M}=0$, relation (41) implies the upper bound
for free energy, which is equal to $\lambda_0$ (\ref{42}).
{}For $\wt{k}>0$, $\wt{M}>0$, this estimate can be improved and,
for $\wt{k}\to k$, $\wt{M}\to M$, coincides with the exact value
of the quantum free energy.

\section{Metastable states and the case of many clusters}

The same procedure can be used for any number of clusters both
in the fermion and boson cases.
Namely, we construct the sypersecondary quantization for this
set of clusters and set the creation and annihilation
operators equal to functions
(i.e., we assume that they commute, and hence we take the
symbols of these operators;
the physicists say: `` we assume that the creation and
annihilation operators are $'$-numbers).

Next, according to this nonlinear problem, we construct a set of
independent functions complete in the corresponding space,
which, in turn, determines the passage to the occupation numbers.
Then we employ the ideology of the Tamm--Dankov method.

The set of such functions is constructed just as above, namely,
the vector system (22) is naturally generalized to the case of
an arbitrary number of clusters and the ratios of matrix
elements similar to (37) are studied for this system.

Expression (37) is the upper bound for the least quantum value
of free energy. Hence it is natural to choose the function system
so that this expression for each function attain its extremum value.
Previously,
this was performed in such a way that equation (38)
is an extremum equation.
{}For an arbitrary number of clusters, the approximate solutions
of the equation for free energy can be constructed by using
a variational method similar to (41). 

Hence, in some sense, this method is similar to the Tamm--Dankov
method, since inequality (42) is used. 
This variational method is especially effective for
systems of third quantized fields and automatically can be
used for such systems.

The supersecondary quantization method can be used in quantum
electrodynamics and quantum chromodynamics,
when the ``charge'' is preserved rather than the number of
particles.
Such an example is studied earlier.
However, in this case we encounter the renormalization problem
and must construct our variational method following the method
proposed by Dyson
[8],
i.e., following the ``new Tamm--Dankov'' method.

To find metastable states corresponding to superfluidity and
superconductivity,
we perform constructions similar to those considered in the
Introduction.
Let $0<\alpha<1$.
We consider the following class of vectors in the space $\cF$:
\begin{equation}
\wt{\Phi}=\sum_{\{n\},\{m\}}B_{\{n\}\{m\}}\Phi_{\{n\},\{m\}},
\end{equation}
where $B_{\{n\}\{m\}}$ depend on the sets of numbers $\{n\}$,
$\{m\}$ so that
$$
B_{\{n\}\{m\}}=0
$$
if at least one of the conditions
\begin{equation}
\sum_{a=1}^{\infty}n_a\leq\alpha M,\qquad
\sum_{c=1}^{\infty}m_c\leq\alpha k
\end{equation}
is not satisfied.
The metastable state corresponding to our function system exists
if the functional
\begin{equation}
\frac{\left(\wt{\Phi},\left(\ov{\wh{\cH}}+\theta\ov{\wh{S}}\right)\wt{\Phi}
\right)}{\left(\wt{\Phi},\ov{\wh{E}}\wt{\Phi}\right)}
\end{equation}
has a minimum on the class of vectors (43) for some value of
$\alpha$.
In the number of cluster types is $\geq2$,
the class of vectors is determined similar to (43)
by using inequalities similar to (47).
And the existence of a metastable state for a larger number of clusters
is determined by the existence of a minimum of the corresponding
functional of type (45)
on the corresponding vector class.

\section{}

We proposed a quantization of entropy and of free energy for
many particles. 

The results can readily be carried over to the case of many fields. 

In particular, we can define
quantum thermodynamics in the case of many fields by generalizing
our operation of quantization over clusters
to the case of third quantization [9] (this operation
was performed for the second quantization).

Let us perform the quantization of thermodynamics in the theory of
many scalar fields. In~[9], a~third quantization of the model
of $N$ scalar fields is presented for the Hamiltonian of the model
of the form
\begin{equation}\label{r1}
\wh{H}_N=\frac12\sum_{a=1}^N\int dx\left(\hp_a(x)\hp_a(x)+\nb\hf_a(x)
\nb\hf_a(x)+m^2\hf_a(x)\hf_a(x)\right)+
\varepsilon\sum_{a,b=1}^N\int dx\hf_a^2(x)\hf_b^2(x),
\end{equation}
where $x\in\bR^3$, $\varepsilon$, and $m$ are given quantities and
$\hp_a(x)$ and $\hf_a(x)$ are operators in the space $\cL_N$
of the symmetric functionals,
$\Psi_N\bigl[\vp_1(\cdot),\dots,\vp_N(\cdot)\bigr]$, of $N$ functions
$\vp_1(\cdot),\dots,\vp_N(\cdot)$,
\begin{eqnarray*}
&&\hf_a(\cdot)\Psi_N\bigl[\vp_1(\cdot),\dots,\vp_N(\cdot)\bigr]=
\vp_a(\cdot)\Psi_N\bigl[\vp_1(\cdot),\dots,\vp_N(\cdot)\bigr],\\
&&\hp_a(\cdot)\Psi_N\bigl[\vp_1(\cdot),\dots,\vp_N(\cdot)\bigr]=
-i\frac{\delta}{\delta \vp_a(\cdot)}
\Psi_N\bigl[\vp_1(\cdot),\dots,\vp_N(\cdot)\bigr].
\end{eqnarray*}

Let us consider the Fock space
$\cL=\oplus_{N=0}^\infty\cL_N$ whose elements are sequences
of symmetric functionals
$\Psi_N\bigl[\vp_1(\cdot),\dots,\vp_N(\cdot)\bigr]$, $N=0,1,\dots$,
and introduce operators of generalized functionals
$\wh{A}^{\pm}\bigl[\vp(\cdot)\bigr]$ on $\cL$ as follows:
\begin{eqnarray}\label{r2}
&&\left(\int D\vp\wh{A}^+\bigl[\vp(\cdot)\bigr]
X\bigl[\vp(\cdot)\bigr]\Psi\right)_k
\bigl[\vp_1(\cdot),\dots,\vp_k(\cdot)\bigr]=\nn\\
&&\qquad=\frac1{\sqrt{k}}\sum_{a=1}^k
X\bigl[\vp_a(\cdot)\bigr]\Psi_{k-1}
\bigl[\vp_1(\cdot),\dots,\vp_{a-1}(\cdot),
\vp_{a+1}(\cdot),\dots,\vp_k(\cdot)\bigr],\\
&&\left(\int D\vp\wh{A}^-\bigl[\vp(\cdot)\bigr]
X^*\bigl[\vp(\cdot)\bigr]\Psi\right)_{k-1}
\bigl[\vp_1(\cdot),\dots,\vp_{k-1}(\cdot)\bigr]=\nn\\
&&\qquad=\sqrt{k}\int D\vp X^*\bigl[\vp(\cdot)\bigr]\Psi_k
\bigl[\vp(\cdot),\vp_1(\cdot),\dots,\vp_{k-1}(\cdot)\bigr].\nn
\end{eqnarray}
To the Hamiltonian of $N$ fields of the form (46), we assign
the following operator
$\wh H$ on the Fock space $\cL$:
$$
(\wh{H}\Psi)_N\bigl[\vp_1(\cdot),\dots,\vp_N(\cdot)\bigr]=
\wh{H}_N\Psi_N\bigl[\vp_1(\cdot),\dots,\vp_N(\cdot)\bigr],
$$
which can be represented as
\begin{eqnarray}\label{r3}
&&\wh{H}=\frac12\int D\vp\wh{A}^+\bigl[\vp(\cdot)\bigr]
\int dx\left(-\frac{\delta^2}{\delta\vp(x)\delta\vp(x)}+\nb\vp(x)\nb\vp(x)+
m^2\vp(x)\vp(x)\right)\wh{A}^-\bigl[\vp(\cdot)\bigr]+\nn\\
&&\qquad+\varepsilon\int\int D\vp D\phi\int dx\vp^2(x)\phi^2(x)
\wh{A}^+\bigl[\vp(\cdot)\bigr]\wh{A}^-\bigl[\vp(\cdot)\bigr]
\wh{A}^+\bigl[\phi(\cdot)\bigr]\wh{A}^-\bigl[\phi(\cdot)\bigr].
\end{eqnarray}
The expression (48) is the triply quantized representation of the
Hamiltonian (46).

Now  we introduce the {\it
cluster\/} spaces $\cM_{N_1,\dots,N_k}$, where $k\geq 1$,
$N_1,\dots,N_k\geq 0$ are integers.
The elements of these spaces are functionals of the form
\begin{eqnarray}\label{r4}
&&\Psi_{N_1,\dots,N_k}\bigl[\vp_1^1(\cdot),j_1^1,\dots,
\vp_{N_1}^1(\cdot),j_{N_1}^1;
\vp_1^2(\cdot),\vp_2^2(\cdot),j_1^2,\dots
\vp_{2N_2-1}^2(\cdot),\vp_{2N_2}^2(\cdot),j_{N_2}^2;\nn\\
&&\qquad\dots;
\vp_1^k(\cdot),\dots,\vp_k^k(\cdot),j_1^k,\dots
\vp_{kN_k-k+1}^k(\cdot),\dots,\vp_{kN_k}^k(\cdot),j_{kN_k}^k\bigl],
\end{eqnarray}
that are symmetric with respect to the permutations of
$$
\vp^l_{pl+1}(\cdot),\dots,\vp^l_{pl+l}(\cdot),j^l_{p+1}, 
\vp^l_{ql+1}(\cdot),\dots,\vp^l_{ql+l}(\cdot),j^l_{q+1}
$$ 
for all $l=1,\dots,k$ and
$p,q=0,\dots,N_l-1$, where $j^l_s$, $s=1,\dots,N_l$, take values
in the set $0,1,\dots,I_l$. We also
introduce the {\it cluster\/} Fock space
$\cM=\oplus_{N_1,\dots,N_k=0}^{\infty}\cM_{N_1,\dots,N_k}$,
which consists of sequences of functionals (49).
Similarly to (47), we can introduce operator generalized
functionals
$\wh{B}^{\pm}_l\bigl[\vp^l_1(\cdot),\dots,\vp^l_l(\cdot),j^l\bigr]$.

We introduce an {\it operator measure\/}
$\wh\rho$, which is an operator on $\cL$ and depends on the
functionals
$B_l\bigl[\vp^l_1(\cdot),\dots,\vp^l_l(\cdot),j^l\bigr]$:
\begin{eqnarray}
&&\wh{\rho}=\exp\left(-\int D\vp\str{2}{\wh{A}^+}\bigl[\vp(\cdot)\bigr]
\str{1}{\wh{A}^-}\bigl[\vp(\cdot)\bigr]\right)\times\nn\\
&&\qquad\times\exp\Bigl(\sum_{l=1}^k\sum_{j^l=0}^{I_l}\int\dots\int
D\vp^l_1D\phi^l_1\dots D\vp^l_lD\phi^l_l\nn\\
&&\qquad \times
B_l^*\bigl[\vp^l_1(\cdot),\dots,\vp^l_l(\cdot),j^l\bigr]
B_l\bigl[\vp^l_1(\cdot),\dots,\vp^l_l(\cdot),j^l\bigr]\times\nn\\
&&\qquad\times
\str{2}{\wh{A}^+}\bigl[\vp^l_1(\cdot)\bigr]
\str{1}{\wh{A}^-}\bigl[\vp^l_1(\cdot)\bigr]
\str{2}{\wh{A}^+}\bigl[\vp^l_l(\cdot)\bigr]
\str{1}{\wh{A}^-}\bigl[\vp^l_l(\cdot)\bigr]\Bigr),
\label{r5}
\end{eqnarray}
where the numbers over the operators stand for the order
of their execution [5].
We omit the arguments $\wh\rho$ in what follows.
{}Furthermore, let
\begin{eqnarray}\label{r6}
&&\cH\left(B^*_1\bigl[\cdot\bigr],B_1\bigl[\cdot\bigr],\dots
B^*_k\bigl[\cdot\bigr],B_k\bigl[\cdot\bigr]\right)=\nn\\
&&\qquad=\exp\left(\sum_{l=1}^k\sum_{j^l=0}^{I_l}\int\dots\int
D\vp^l_1\dots D\vp^l_l
B^*_l\bigl[\vp^l_1(\cdot),\dots,\vp^l_l(\cdot)\bigr]
B_l\bigl[\vp^l_1(\cdot),\dots,\vp^l_l(\cdot)\bigr]\right)\times\nn\\
&&\qquad\times\Sp\left(\wh{\rho}\wh{H}\right).
\end{eqnarray}

By the {\it cluster
representation} of the triply quantized Hamiltonian (\ref{r3})
we mean the following operator on $\cM$:
\begin{equation}\label{r7}
\wh{\cH}=\cH\left(\str{2}{\wh{B}^+_1}\bigl[\cdot\bigr],
\str{1}{\wh{B}_1^-}\bigl[\cdot\bigr],\dots
\str{2}{\wh{B}^+_k}\bigl[\cdot\bigr],
\str{1}{\wh{B}_k^-}\bigl[\cdot\bigr]\right).
\end{equation}

We introduce the operator of {\it
entropy\/} of many fields on $\cM$ by the relation
\begin{equation}\label{r8}
\wh{\cS}=\cS_{reg}\left(\str{2}{\wh{B}^+_1}\bigl[\cdot\bigr],
\str{1}{\wh{B}_1^-}\bigl[\cdot\bigr],\dots
\str{2}{\wh{B}^+_k}\bigl[\cdot\bigr],
\str{1}{\wh{B}_k^-}\bigl[\cdot\bigr]\right),
\end{equation}
where $\cS_{reg}$ is the regularized functional
\begin{equation}\label{r9}
\cS=\Sp\left(\wh{\rho}\ln\left(\frac{\wh{\rho}}{\Sp\wh{\rho}}\right)\right).
\end{equation}

\hbox{}

{\bf Definition}. The numbers $\lambda$ such that
\begin{equation}\label{r10}
\left(\wh{\cH}+\theta\wh{\cS}\right)\Phi=\lambda\Phi,\qquad\Phi\in\cM,\,
\Phi\ne0,
\end{equation}
are called {\it quantum values of free energy\/} of many fields
under the temperature~$\theta$.

\hbox{}

In the general case of a system of many fields, the quantization of
entropy and free energy can be carried out just as was done
in the present note.

\section{Supersecondary quantization and quantization of entropy
with preserved charge}

Earlier the supersecondary quantization
and quantization of thermodynamics was introduced in the case
in which the secondary quantized Hamiltonian of the system
commutes with the operator of the number of particles.
In this section the secondary quantization and quantization of
thermodynamics with number of cluster types equal to~$2$
is generalized to the case of charge preservation.

Let $n_1,n_2,m\geq 0$ be integers. Consider the functions
\begin{equation}\label{u1}
\Phi_{n,m_1,m_2}(x_1,j_1,\dots,x_{n_1},j_{n_1},y_1,i_1,\dots,y_{n_2},i_{n_2},
z_1,w_1,\dots,z_m,w_m)
\end{equation}
that are symmetric with respect to permutations of the pairs
$x_p,j_p$ and $x_q,j_q$,
with respect to permutations of the pairs
$y_r,i_r$ and $y_s,i_s$,
and with respect to permutations of the pairs
$z_l,w_l$ and $z_t,w_t$, where the variables $x_p$, $y_r$,
$z_t$, $w_l$ take values on the three-dimensional torus $\bT^3$,
and the variables $j_p$ and $i_r$
attain values $1,2,\dots$ and are called {\it numbers}.
The Hilbert space of functions (56) with the norm
\begin{eqnarray}
&&\|\Phi_{n_1,n_2,m}\|^2=\nn\\
&&=\sum_{j_1=1}^{\infty}\dots
\sum_{j_{n_1}=1}^{\infty}\dots
\sum_{i_1=1}^{\infty}\dots
\sum_{i_{n_2}=1}^{\infty}\int\!\dots\!\int dx_1\dots dx_{n_1}
dy_1\dots dy_{n_2}
dz_1dw_1\dots dz_mdw_m\times\nn\\
&&\qquad\times
|\Phi_{n,m_1,m_2}(x_1,j_1,\dots,x_{n_1},j_{n_1},y_1,i_1,\dots,y_{n_2},i_{n_2},
z_1,w_1,\dots,z_m,w_m)|^2
\label{u2}
\end{eqnarray}
we denote by $\cF_{n_1,n_2,m}$.
Now we introduce the {\it cluster\/} space $\cF$:
\begin{equation}\label{u3}
\cF=\bigoplus_{n_1,n_2,m=0}^{\infty}\cF_{n_1,n_2,m},
\end{equation}
whose elements are sets of functions (56):
\begin{eqnarray*}
&&\Phi=\{\Phi_{n_1,n_2,m}\},\qquad n_1,n_2,m=0,1,\dots,\\
&&\|\Phi\|^2=\sum_{n_1,n_2,m=0}^{\infty}\|\Phi_{n_1,n_2,m}\|^2.
\end{eqnarray*}

The space $\cF$ is a special case of the Fock space
in which we introduce the creation and annihilation operators
$\wh{b}^{\pm}_1(x,j)$, $\wh{b}^{\pm}_2(y,i)$, $\wh{B}^{\pm}(z,w)$:
\begin{eqnarray*}
&&(\wh{b}^+_1(x,j)\Phi)_{n_1,n_2,m}
(x_1,j_1,\dots,x_{n_1},j_{n_1},y_1,i_1,\dots,y_{n_2},i_{n_2},
z_1,w_1,\dots,z_m,w_m)=\\
&&\qquad=\frac{1}{\sqrt{n_1}}\sum_{l=1}^{n_1}\delta_{jj_l}\delta(x-x_l)
\Phi_{n_1-1,n_2,m}
(x_1,j_1,\dots,x_{l-1},j_{l-1}
,x_{l+1},j_{l+1},\dots,x_{n_1},j_{n_1},\\
&&\qquad y_1,i_1,\dots,y_{n_2},i_{n_2},
z_1,w_1,\dots,z_m,w_m),\\
&&(\wh{b}^-_1(x,j)\Phi)_{n_1,n_2,m}
(x_1,j_1,\dots,x_{n_1},j_{n_1},y_1,i_1,\dots,y_{n_2},i_{n_2},
z_1,w_1,\dots,z_m,w_m)=\\
&&\qquad=\sqrt{n_1+1}
\Phi_{n_1+1,n_2,m}
(x,j,x_1,j_1,\dots,x_{n_1},j_{n_1},y_1,i_1,\dots,y_{n_2},i_{n_2},
z_1,w_1,\dots,z_m,w_m),\\
&&(\wh{b}^+_2(y,i)\Phi)_{n_1,n_2,m}
(x_1,j_1,\dots,x_{n_1},j_{n_1},y_1,i_1,\dots,y_{n_2},i_{n_2},
z_1,w_1,\dots,z_m,w_m)=\\
&&\qquad=\frac{1}{\sqrt{n_2}}\sum_{l=1}^{n_2}\delta_{ii_l}\delta(y-y_l)
\Phi_{n_1,n_2-1,m}
(x_1,j_1,\dots,x_{n_1},j_{n_1},y_1,i_1,\dots,
y_{l-1},i_{l-1},\\
&&\qquad y_{l+1},i_{l+1},\dots,
y_{n_2},i_{n_2},
z_1,w_1,\dots,z_m,w_m),\\
&&(\wh{b}^-_2(y,i)\Phi)_{n_1,n_2,m}
(x_1,j_1,\dots,x_{n_1},j_{n_1},y_1,i_1,\dots,y_{n_2},i_{n_2},
z_1,w_1,\dots,z_m,w_m)=\\
&&\qquad=\sqrt{n_2+1}
\Phi_{n_1,n_2+1,m}
(j_1,\dots,x_{n_1},j_{n_1},y,i,y_1,i_1,\dots,y_{n_2},i_{n_2},
z_1,w_1,\dots,z_m,w_m),\\
&&(\wh{B}^+(z,w)\Phi)_{n_1,n_2,m}
(x_1,j_1,\dots,x_{n_1},j_{n_1},y_1,i_1,\dots,y_{n_2},i_{n_2},
z_1,w_1,\dots,z_m,w_m)=\\
&&\qquad=\frac{1}{\sqrt{m}}\sum_{l=1}^{m}\delta(z-z_l)\delta(w-w_l)
\Phi_{n_1,n_2,m-1}
(x_1,j_1,\dots,x_{n_1},j_{n_1},y_1,i_1,\dots,
y_{n_2},i_{n_2},\\
&&\qquad z_1,w_1,\dots,z_{l-1},w_{l-1},z_{l+1},w_{l+1},\dots,
z_m,w_m),\\
&&(\wh{B}^-(z,w)\Phi)_{n_1,n_2,m}
(x_1,j_1,\dots,x_{n_1},j_{n_1},y_1,i_1,\dots,y_{n_2},i_{n_2},
z_1,w_1,\dots,z_m,w_m)=\\
&&\qquad=\sqrt{m+1}
\Phi_{n_1,n_2,m+1}
(x_1,j_1,\dots,x_{n_1},j_{n_1},y_1,i_1,\dots,y_{n_2},i_{n_2},
z,w,z_1,w_1,\dots,z_m,w_m),
\end{eqnarray*}
where $\delta_{jj'}$ is the Kronecker delta.
The operators $\wh{b}^+_1(x,j)$ and $\wh{b}^-_1(x,j)$
will be called the {\it creation and annihilation
operators for charge}~$1$,
the operators $\wh{b}^+_2(x,j)$ and $\wh{b}^-_2(x,j)$
will be called the {\it creation and annihilation
operators for charge}~$-1$,
and the operators $\wh{'}^+(x,j)$ and $\wh{b}^-_1(x,j)$
will be called the {\it creation and annihilation
operators for charge}~$0$.
We consider the subspace of the space $\cF_{n_1,n_2,m}$,
which consists of functions (\ref{u1})
symmetric with respect to permutations of the variables
$x_l$ and $x_p$,
symmetric with respect to permutations of the variables
$x_l$ and $z_r$,
symmetric with respect to permutations of the variables
$y_q$ and $y_s$,
and symmetric with respect to permutations of the variables
$y_q$ and $w_t$
and denote this subspace by $\cF_{n_1,n_2,m}^B$.
The following subspace of $\cF$:
\begin{equation}\label{u4}
\cF^B=\bigoplus_{n_1,n_2,m=0}^{\infty}\cF_{n_1,n_2,m}^B
\end{equation}
will be called the {\it boson subspace of the cluster space}.

Now we consider the ``usual'' Fock space $\cH$ of boson systems
with charges $\pm1$, which consists of sets of the functions
$\phi=\{\phi_{q_1,q_2}(x_1,\dots,x_{q_1},y_1,\dots,y_{q_2})\}$, $q_1,q_2=0,1,
\dots$ symmetric with respect to permutations of the variables
$x_l$ and $x_p$
and symmetric with respect to permutations of the variables
$y_k$ and $y_r$. In this space
the creation and annihilation operators $\wh{\psi}^{\pm}_1(x)$
for bosons with charge~$1$
and the creation and annihilation operators
$\wh{\psi}^{\pm}_1(y)$ for bosons with charge~$-1$
are defined as
\begin{eqnarray*}
&&(\wh{\psi}^+_1(x)\phi)_{q_1,q_2}(x_1,\dots,x_{q_1},y_1,\dots,y_{q_2})=\\
&&\qquad=\frac1{\sqrt{q_1}}\sum_{l=1}^{q_1}\delta(x-x_l)
\phi_{q_1-1,q_2}(x_1,\dots,x_{l-1},x_{l+1},\dots,
x_{q_1},y_1,\dots,y_{q_2}),\\
&&(\wh{\psi}^-_1(x)\phi)_{q_1,q_2}(x_1,\dots,x_{q_1},y_1,\dots,y_{q_2})=\\
&&\qquad=\sqrt{q_1+1}\phi_{q_1+1,q_2}(x,x_1,\dots,
x_{q_1},y_1,\dots,y_{q_2}),\\
&&(\wh{\psi}^+_2(y)\phi)_{q_1,q_2}(x_1,\dots,x_{q_1},y_1,\dots,y_{q_2})=\\
&&\qquad=\frac1{\sqrt{q_2}}\sum_{l=1}^{q_1}\delta(y-y_l)
\phi_{q_1,q_2-1}(x_1,\dots,x_{q_1},y_1,\dots,y_{l-1},y_{l+1},\dots,y_{q_2}),\\
&&(\wh{\psi}^-_2(y)\phi)_{q_1,q_2}(x_1,\dots,x_{q_1},y_1,\dots,y_{q_2})=\\
&&\qquad=\sqrt{q_2+1}\phi_{q_1,q_2+1}(x_1,\dots,x_{q_1},y,y_1,\dots,y_{q_2}).
\end{eqnarray*}
An operator in $\cH$ of the form
\begin{equation}\label{u5}
\wh{Q}=\int dx\,\wh{\psi}^+_1(x)\wh{\psi}^-_1(x)
-\int dx\,\wh{\psi}^+_2(x)\wh{\psi}^-_2(x)
\end{equation}
is called the {\it charge operator} $\wh{Q}$.
Let $\wh{A}$ be an operator in the space $\cH$
commuting with the charge operator~$\wh{Q}$.
The {\it supersecondary quantized operator\/}$\ov{\wh{A}}$
is the following operator in $\cF$:
\begin{equation}\label{u6}
\ov{\wh{A}}=A\bigl[\str{2}{\wh{b}^+_1}(x,j),\str{1}{\wh{b}^-_1}(x,j)
,\str{2}{\wh{b}^+_2}(y,i),\str{1}{\wh{b}^-_2}(y,i)
,\str{2}{\wh{B}^+}(z,w),\str{1}{\wh{B}^-}(z,w)\bigl],
\end{equation}
where the numbers over the operators stand for order of their
action and the functional
$A\bigl[b^*_1(\cdot),b_1(\cdot),b^*_2(\cdot),b_2(\cdot),
B^*(\cdot),B(\cdot)\bigl]$ has the form
\begin{eqnarray}
&&A\bigl[b^*_1(\cdot),b_1(\cdot),b^*_2(\cdot),b_2(\cdot)
,B^*(\cdot),B(\cdot)\bigl]=\Sp\left(\wh{A}\wh{\rho}
\bigl[b^*_1(\cdot),b_1(\cdot),b^*_2(\cdot),b_2(\cdot)
,B^*(\cdot),B(\cdot)\bigl]\right)\times\nn\\
&&\qquad\times\exp\left(-\sum_{j=1}^{\infty}
\int dx\,b^*_1(x,j)b_1(x,j)-\sum_{i=1}^{\infty}\int
dy\,b^*_2(y,i)b_2(y,i)\right),
\label{u7}
\end{eqnarray}
where $\wh{\rho}\bigl[b^*_1(\cdot),b_1(\cdot),b^*_2(\cdot),
b_2(\cdot),B^*(\cdot),B(\cdot)\bigr]$ is an operator valued
functional ranging in the set of operators in the space $\cH$
and having the form
\begin{eqnarray}
&&\wh{\rho}\bigl[b^*_1(\cdot),b_1(\cdot),b^*_2(\cdot),b_2(\cdot)
,B^*(\cdot),B(\cdot)\bigr]=\nn\\
&&\qquad=
\exp\left(\frac12\int\!\!\int dz_1dw_1\,B(z_1,w_1)\str{2}{\wh{\psi}^+_1}(z_1)
\str{2}{\wh{\psi}^+_2}(w_1)\right)\times\nn\\
&&\qquad\times
\exp\left(\frac12\int\!\!\int dz_2dw_2\,B^*(z_2,w_2)
\str{1}{\wh{\psi}^-_1}(z_2)
\str{1}{\wh{\psi}^-}(w_2)\right)\times\nn\\
&&\qquad\times
\exp\left(\sum_{j=1}^{\infty}\int\!\!\int dx_1dx_2\,b_1(x_1,j)b^*_1(x_2,j)
\str{2}{\wh{\psi}^+_1}(x_1)
\str{1}{\wh{\psi}^-_1}(x_2)\right)\times\nn\\
&&\qquad\times
\exp\left(\sum_{i=1}^{\infty}\int\!\!\int dy_1dy_2\,b_2(y_1,i)b^*_2(y_2,i)
\str{2}{\wh{\psi}^+_2}(y_1)
\str{1}{\wh{\psi}^-_2}(y_2)\right)\times\nn\\
&&\qquad\times
\exp\left(-\int dx\str{2}{\wh{\psi}^+_1}(x)\str{1}{\wh{\psi}^-_1}(x)
-\int dy\str{2}{\wh{\psi}^+_2}(y)\str{1}{\wh{\psi}^-_2}(y)\right).
\label{u8}
\end{eqnarray}
To each element $\Phi$ of the boson subspace $\cF^B$ of the space $\cF$,
we bijectively assign the set
$\{\phi (j_1,\dots,j_{q_1},i_1,\dots,i_{q_2})\}$, $q_1,q_2=0,1,\dots$,
with values in the space $\cH$ of the functions of the numbers
\begin{eqnarray}
&&(\phi(j_1,\dots,j_{q_1},i_1,\dots,i_{q_2}))_{n_1,n_2}
(x_1,\dots,x_{n_1},y_1,\dots,y_{n_2})=\nn\\
&&\qquad=
\delta_{q_1-n_1,q_2-n_2}
\theta(n_1-q_1)\Phi_{q_1,q_2,n_1-q_1}(x_1,j_1,\dots,x_{q_1},j_{q_1},
y_1,i_1,\dots,y_{q_2},i_{q_2},\nn\\
&&\qquad\qquad x_{q_1+1},y_{q_2+1},\dots,x_{n_1},y_{n_2}),
\label{u9}
\end{eqnarray}
where $\theta(\cdot)$ is the Heaviside theta function.
Hence, to the secondary quantized operator $\wh{A}$,
there naturally corresponds an operator $\wt{\wh{A}}$ in the
space $\cF^B$ such that
\begin{eqnarray}
&&(\wt{\wh{A}}\Phi)_{q_1,q_2,m}(x_1,j_1,\dots,x_{q_1},j_{q_1},y_1,i_1,\dots,
y_{q_2},i_{q_2},x_{q_1+1},y_{q_2+1},\dots,x_{q_1+m},y_{q_2+m})=\nn\\
&&\qquad=(\wh{A}
\phi(j_1,\dots,j_{q_1},i_1,\dots,i_{q_2}))_{q_1+m,q_2+m}
(x_1,\dots,x_{q_1+m},y_1,\dots,y_{q_2+m}).
\label{u10}
\end{eqnarray}
We have the following theorem.

\hbox{}

{\bf Theorem}. {\it The projection of the supersecondary
quantized operator $\ov{\wh{A}}$ on the boson subspace
is equal to $\wt{\wh{A}}$.}

\hbox{}

We introduce an operator of supersecondary quantized entropy
$\ov{\wh{S}}$
for systems with preserved charge
\begin{equation}\label{u11}
\ov{\wh{S}}=S_{reg}\bigl[\str{2}{\wh{b}^+_1}(x,j),\str{1}{\wh{b}^-_1}(x,j)
,\str{2}{\wh{b}^+_2}(y,i),\str{1}{\wh{b}^-_2}(y,i)
,\str{2}{\wh{B}^+}(z,w),\str{1}{\wh{B}^-}(z,w)\bigl],
\end{equation}
where the functional $S_{reg}\bigl[b^*_1(\cdot),b_1(\cdot),
b^*_2(\cdot),b_2(\cdot),B^*(\cdot),B(\cdot)\bigl]$
is equal to the regularized functional
\begin{eqnarray}
&&S\bigl[b^*_1(\cdot),b_1(\cdot),b^*_2(\cdot),b_2(\cdot)
,B^*(\cdot),B(\cdot)\bigl]=\nn\\
&&\quad=\Sp\left(\wh{\rho}
\bigl[b^*_1(\cdot),b_1(\cdot),b^*_2(\cdot),b_2(\cdot)
,B^*(\cdot),B(\cdot)\bigl]
\frac{
\wh{\rho}
\bigl[b^*_1(\cdot),b_1(\cdot),b^*_2(\cdot),b_2(\cdot)
,B^*(\cdot),B(\cdot)\bigl]}{
\Sp\wh{\rho}\bigl[b^*_1(\cdot),b_1(\cdot),b^*_2(\cdot),b_2(\cdot)
,B^*(\cdot),B(\cdot)\bigl]}\right)\times\nn\\
&&\qquad\times\exp\left(-\sum_{j=1}^{\infty}
\int dx\,b^*_1(x,j)b_1(x,j)-\sum_{i=1}^{\infty}\int
dy\,b^*_2(y,i)b_2(y,i)\right).
\label{u12}
\end{eqnarray}

\end{document}